\documentclass{aastex6}

\usepackage{amsmath}
\usepackage{graphicx}
\usepackage{hyperref}
\usepackage{ctable}
\usepackage{float}

\newcommand{\refeq}[1]{Eq.~(\ref{eq:#1})}          
\newcommand{\refeqs}[2]{Eqs.~(\ref{eq:#1})--(\ref{eq:#2})}          
\newcommand{\reffig}[1]{Figure~\ref{fig:#1}}          
\newcommand{\refsec}[1]{Section~\ref{sec:#1}}

\newcommand{\reftab}[1]{Table~\ref{tab:#1}}

\def\bfz{\boldsymbol{z}}
\def\bfn{\boldsymbol{n}}

\def\bfe{\boldsymbol{e}}
\def\bfr{\boldsymbol{r}}
\def\bfx{\boldsymbol{x}}
\def\bfy{\boldsymbol{y}}

\def\bfd{\boldsymbol{d}}

\def\bfV{\boldsymbol{V}}

\def\bfd{\boldsymbol{d}}

\def\bfalp{\boldsymbol{\alpha}}
\def\b{\boldsymbol}

\def\d{\mathrm{d}}

\newcommand{\be}{\begin{equation}}
\newcommand{\ee}{\end{equation}}
\newcommand{\ba}{\begin{eqnarray}}
\newcommand{\ea}{\end{eqnarray}}
\newcommand{\en}{\nonumber\\}

\begin{document}


\title{Probing motion of fast radio burst sources by timing strongly lensed repeaters}



\author{Liang Dai\footnote{NASA Einstein Fellow.}}
\affil{School of Natural Sciences, Institute for Advanced Study\\
1 Einstein Drive \\
Princeton, NJ 08540, USA}

\author{Wenbin Lu}
\affil{Department of Astronomy, The University of Texas at Austin\\
2515 Speedway, Stop C1400\\
Austin, TX 78712, USA}




\begin{abstract}

Given the possible repetitive nature of fast radio bursts (FRBs), their cosmological origin, and their high occurrence, detection of strongly lensed sources due to intervening galaxy lenses is possible with forthcoming radio surveys. We show that if multiple images of a repeating source are resolved with VLBI, using a method independent of lens modeling, accurate timing could reveal non-uniform motion, either physical or apparent, of the emission spot. This can probe the physical nature of FRBs and their surrounding environments, constraining scenarios including orbital motion around a stellar companion if FRBs require a compact star in a special system, and jet-medium interactions for which the location of the emission spot may randomly vary. The high timing precision possible for FRBs ($\sim {\rm ms}$) compared to the typical time delays between images in galaxy lensing ($\gtrsim 10\, {\rm days}$) enables the measurement of tiny fractional changes in the delays ($\sim 10^{-9}$), and hence the detection of time-delay variations induced by relative motions between the source, the lens, and the Earth. We show that uniform cosmic peculiar velocities only cause the delay time to drift linearly, and that the effect from the Earth's orbital motion can be accurately subtracted, thus enabling a search for non-trivial source motion. For a timing accuracy of $\sim 1\,$ms and a repetition rate (of detected bursts) $\sim 0.05$ per day of a single FRB source, non-uniform displacement $\gtrsim 0.1 - 1\,$AU of the emission spot perpendicular to the line of sight is detectable if repetitions are seen over a period of hundreds of days.

\end{abstract}




\section{introduction}
\label{sec:intro}

Fast radio bursts (FRBs) are bright
transients discovered at $\sim\rm GHz$ frequencies with millisecond durations
\citep{2007Sci...318..777L, 2013Sci...341...53T}. Their dispersion
measures (DMs), which measure the free electron column density along the line of sight toward the source
${\rm DM} = \int n_{\rm e}\,\d l$, exceed the contribution from the Galactic
interstellar medium (ISM) by typically an order of magnitude. If intergalactic medium (IGM) accounts for most of the DM excess, these bursts must have travelled across
cosmological distances. Recently, the astrophysical nature of these radio transients has become one of the most intriguing mysteries in astronomy.

A major breakthrough was the serendipitous observation that one of the known bursts,
FRB 121102, is sporadically repeating \citep{
 2016Natur.531..202S}. Very recently, this repeater has been successfully localized to 
sub-arcsecond resolution thanks to the Jansky Very Large
Array and the European VLBI Network \citep{2017Natur.541...58C,
  2017ApJ...834L...8M}. It was found to be in association with
a dwarf star-forming host galaxy at redshift $z = 0.19$
\citep{2017ApJ...834L...7T}, thus confirming its cosmological
origin. FRB 121102 was the first and is to date the only FRB discovered by
the Arecibo Observatory \citep{2014ApJ...790..101S}. Despite dedicated follow-up
monitoring \citep[e.g.][]{2015ApJ...799L...5R, 2015MNRAS.454..457P},
none of the other known FRBs (mostly found by the Parkes telescope) has been observed
to repeat. This may suggest two physically distinct classes: non-repeating scenarios invoking cataclysmic processes (e.g. \cite{Hansen:2000am, Piro:2012rq, Totani:2013lia, Kashiyama:2013gza, falcke2014fast, fuller2015dark, 2014ApJ...780L..21Z}); repeating mechanisms in which the source can last for long (e.g. \cite{kulkarni2014giant, Lyubarsky:2014jta, Cordes:2015fua, Katz:2015ltv, 2016ApJ...829...27D, connor2016non, 2016MNRAS.462..941L, Zhang:2017zse, 2017MNRAS.468.2726K}). On the other hand, this may also be an observational bias due to the lower sensitivity and the coarser localization of Parkes relative to Arecibo\footnote{If FRB 121102 had a location error similar to those of the typical FRBs found by Parkes and it were
  to be followed up by Parkes, the true location may
  fall onto the low-sensitivity gaps between beams and perhaps none of the
  subsequent bursts could have been detected.}. Indeed, observations so far are statistically
consistent with all FRBs being repeaters with a repetition frequency and a peak flux
distribution similar to those of FRB 121102 \citep{2016MNRAS.461L.122L}. 

The estimation for the all-sky FRB rate above $\sim 2\rm\ Jy\ ms$ is high $\sim10^3-10^4\,{\rm day}^{-1}
$ \citep{2013Sci...341...53T, 2016MNRAS.460L..30C}. Since FRBs can be visible out to cosmological distances $z \sim 1$, about $\sim {\rm few} \times 10^{-4}$ of them \citep{Hilbert:2007jd} are expected to be
gravitationally lensed by intervening galaxies
\citep{2014SCPMA..57.1390L,Dai:2016igl}. In fact, the observed strong lensing fraction should be higher than the prior probability due to magnification bias. Dominated by lens mass scales $\gtrsim 10^{10}\,M_\odot$, this creates image multiplets separated by arcseconds, with mutual time delays typically on the order of weeks to months. In particular, a repeating source, if lensed, will result in a set of multiple bursts for each repetition. With the prospect that forthcoming large-scale radio surveys,
including UTMOST \citep{2017arXiv170310173C}, HIRAX
\citep{2016SPIE.9906E..5XN}, CHIME \citep{2014SPIE.9145E..22B}, and later on SKA1 \citep{Macquart:2015uea},
will have the capacity to find $\sim10^2 - 10^4$ FRBs per year,
the interesting situation of a strongly lensed FRB repeater becomes worthy of consideration. 

Since survey telescopes will be efficient at detecting a large number of
sources, one realistic way to find strongly lensed events is to look
up the catalog for special ones. Although typical survey telescopes
are not able to spatially resolve multiple images, it is still
possible to distinguish a lensed repeater from unlensed ones. Image
multiplets should have coincidental locations on the sky up to
localization error, and are expected to have similar but not identical DMs. Depending on what frequencies to observe, lensed bursts may be significantly scatter-broadened compared to ordinary ones, due to ISM in the lens galaxy. Furthermore, a series of image multiplets from the same source will exhibit a fixed pattern in their mutual time delays, appearing over and over again as we detect its repetitions one after another. Recognizing such a temporal pattern along a certain line of sight will uncover a strongly lensed repeater.  

The discovery of strongly lensed repeaters should justify the use of more expensive observational resources in order to study them in greater detail. It will then be desirable to capture more repetitions in deep VLBI observations on sub-arcsecond angular scales, which will resolve multiple images and identify the host and the lens. Compared to other variable sources subject to strong lensing at cosmological distances, such as supernovae (SNe) and quasars, FRBs are unique because timing accuracy on the order of milliseconds is achievable, owing to their extremely narrow (de-dispersed) widths. This leads to the question of what astrophysics can we potentially learn by exploiting this high level of precision? Previously, it has been suggested that microlensing time delay can probe compact mass clumps along the line of sight \citep{Munoz:2016tmg}. In this work, we explore the possibility that galaxy-lensing time delays can probe non-uniform motion of the source on $\sim$AU scales, and hence constrain its astrophysical nature and its surrounding environment.

From one repetition to the next, the lensing delay time between a pair of images varies due to the motion of the source~\citep{1999ApJ...519L..31Y, 2002MNRAS.334..905G}, the lens, and the observer. Velocities that are quasi-constant over the observational time span ($\sim$ yrs) can induce a linear drift in the delay. Those include cosmological peculiar velocities of the host galaxy, the lens galaxy, and the Milky Way, as well as the source's large-scale motion within its host, and the Solar System's motion within the Milky Way. Numerically large, those are not predictable for individual cases, but their statistics can be used to measure cosmological parameters and structure formation~\citep{1996ApJ...473..610K}. For this work, we will account for their effects but assume they are not of main interest here.

By contrast, motions that are non-uniform over the observational timescale leave non-trivial signatures in the time delay on top of linear drifts. The Earth's orbital motion generates a sinusoidal perturbation $\lesssim 10^3\,{\rm s}$ to the delay time, whose amplitude and phase provide information to indirectly localize the source to $\sim 10''$, which may then facilitate interferometric follow-ups. Furthermore, as we will show, if multiple images are well resolved with VLBI and the host and lens redshifts are obtained, then the effect of the Earth's orbital motion can be subtracted down to millisecond accuracy. Any additional non-trivial variation in the delay time probes non-uniform source motion transverse to the line of sight. As we will show, this method has the advantage that it does not require lens modeling.

This powerful method, if realized, will constrain the astrophysical nature of repeating FRBs. If spots of coherent radio emission wander around across a transverse region of size $\gtrsim\,1$AU, stochastic variations will be imprinted in the delay time. This may occur when mini-jets, produced through dissipation of magnetic energy inside a larger but slow jet \citep{giannios2009fast}, collide with clouds in the ambient medium \citep{Romero:2015nec}. In such scenario, radiation comes from slightly different locations from one burst to another. Non-detection of such behavior constrains the compactness of the source system, narrowing down possible astrophysical scenarios. 

Many FRB models are based on young neutron stars \citep[NSs, see][for
a brief review]{2016MPLA...3130013K}, in which case highly beamed \citep{Lyubarsky:2014jta} or near-surface
radio emission \citep{2017MNRAS.468.2726K} is not expected to cause
detectable variations in the delay time. On the other hand, a NS with
a stellar companion will imprint an effect due to its orbital
motion. Core-collapse SNe in a binary system are more likely to give
birth to isolated NSs because the sudden mass loss and NS natal kick
tend to unbind the system. However, when the kick velocity is small
($\lesssim100\rm\ km/s$) and the NS is kicked in the opposite
direction of the pre-supernova orbital motion, the binary system may
survive \citep[and roughly a few percent of them do
survive,][]{1983ApJ...267..322H}. In this case, the eccentric orbit of
a young system can have a semi-major axis large enough to produce
noticeable signatures. For example, the binary pulsar PSR B1820-11, a relatively young 
neutron star at an age $\sim 3\,{\rm Myr}$, has an eccentric orbit with 
$e \simeq 0.8$ \citep{lyne1989psr} and a large semi-major axis $\sim 1.3\,{\rm AU}$
\footnote{Australia Telescope National Facility Pulsar Group, 2004, ``ATNF Pulsar Catalogue'', \url{http://www.atnf.csiro.au/research/pulsar/psrcat/}}. 
Many known binary pulsars with main-sequence companions are found to have $e \gtrsim 0.6$ 
and orbital semi-major axes $ \gtrsim 0.1\,{\rm AU}$, while binary pulsars with white-dwarf 
companions have low eccentricities ($e \ll 0.1$) and a significant fraction of them have wide orbits 
on AU scales (see e.g. \cite{Tauris:2012jp}). If the birth rate for repeaters 
is much lower than that for core-collapse SNe, it may be that FRBs require special
conditions for the source system \citep{2016MNRAS.461L.122L}, which
provides further motivation to probe possible source motion. Moreover,
other FRB models that predict the progenitor to have non-uniform motion
could be testable with lensing time delay. For instance, the repeater
FRB 121102 has been attributed to a NS traveling through an asteroid
belt of another star \citep{2016ApJ...829...27D}, or it is subject to
plasma lensing by density inhomogeneities in the host
galaxy \citep{2017arXiv170306580C}. In both cases, the source position
may have large non-uniform transverse motion $\gtrsim 1\rm\ AU$
(either physical or apparent) detectable with the method we propose
here. Given our limited understanding of FRBs to date, it is
worthwhile to explore the scientific potential of this novel
observational method. 

We organize this paper as follows. In \refsec{order}, we first carry out order-of-magnitude estimates on motion-induced
variations in lensing time delays, offering useful intuition on the relevant physical scales of the problem. Rigorous derivations then follow in \refsec{timedelay}. Then, in \refsec{simulation}, we construct a hypothetical strong-lensing event and describe the
procedure of simulating mock time-delay measurements. This will serve as a realistic example in later sections for numerically assessing the accuracy of time-delay measurement and for verifying the back-of-the-envelope estimates in \refsec{order}. In \refsec{localize}, we study indirect source localization using time delay variation between spatially unresolved multiple images. In
\refsec{source}, we study how non-uniform source motion can be
measured if multiple images are resolved with very-long-baseline
interferometry (VLBI). In
\refsec{discussion}, we discuss how microlensing and scattering might
broaden the pulse and affect timing accuracy. Final remarks will be
made in \refsec{concl}.

\section{Order-of-magnitude estimate}
\label{sec:order}

Before delving into detailed calculations, we first seek intuition by estimating the order of magnitude for the relevant physics scales. For simplicity, redshift factors are neglected. They will be easily recovered later. Since we mainly consider galactic-scale halos $M_h \approx 10^{12} - 10^{13}\,M_\odot$ as the most probable intervening lenses, with typical gravitational radii much longer than the radio wavelength, the language of ray optics is suitable.

Imagine an FRB source located at a typical cosmological distance $D \sim 1\,{\rm Gpc}$ away from the Earth, for which the optical depth to strong lensing by intervening galaxies is small but non-negligible. Assume a lens galaxy between the source and the observer. Typically, the source-lens distance and the lens-observer distance are both of order $D$.

Strong lensing splits each burst from the source into several images. They have typical angular separations $\Delta\theta \sim 1''$, which is not resolvable by single-dish telescopes or short-baseline arrays, but is within the reach of VLBI. They have mutual delays in the time of arrival on the order of $\Delta T \sim 0.01 -1\,{\rm yr}$. For FRBs, the typical (de-dispersed) burst width is remarkably short even after scatter broadening, which enables time of arrival to be measured to an accuracy of $\delta T \sim {\rm ms} \ll \Delta T$. 

If a lensed source sporadically repeats, each repetition produces a set of multiple images. The time delay between a given pair of images, at zeroth order, is the same for all repetitions. However, the agreement is imperfect due to relative motions between the source, the lens, and the Earth, both perpendicular to the line of sight and parallel to the line of sight. Those alter the length of the optical path each burst has to travel before reaching the telescope .

\subsection{Motion of the Earth}

It is convenient to derive various kinematic effects in the rest frame of the Solar-System barycenter. This can be treated as an inertial frame to good approximation, since the acceleration of the Solar System within the Milky Way is negligible over the observational time span. The orbital motion of the Earth in this frame is known to high precision, and hence perturbation to the time delay induced by Earth's motion is predictable.

Due to the line-of-sight component of the Earth's orbital motion, for each burst from the source, multiple images reach the Earth at different times. If the Earth recedes from the source at a velocity $v_{\oplus \parallel}$, to linear order radio wave of a later image travels an additional distance $v_{\oplus \parallel}\,\Delta T$ relative to that of an earlier image, and therefore the mutual time delay changes by $\sim (v_{\oplus \parallel}/c)\,\Delta T$ compared to the case without line-of-sight motion. This causes a measurable imprint between repetitions because $v_{\oplus \parallel}$ varies with time. Indeed, a constant $v_{\oplus  \parallel}$ would not be distinguishable from the delay caused by stationary lensing. The time delay varies annually by an amount
\ba
\label{eq:orbitalLoS}
30\,{\rm km\,s}^{-1} \times \Delta T / c \approx 300\,{\rm s}\,\left( \frac{\Delta T}{0.1\,{\rm yr}} \right),
\ea
up to factors dependent on the orientation of the line of sight with
respect to the Earth's orbital plane, and the maximum time-delay
variation is $2\,{\rm AU}/c \approx 1000\,{\rm s}$. This is much larger than the timing resolution $\delta T \sim {\rm ms}$.

The Earth's orbital motion perpendicular to the {\it direction of wave propagation} (which is slightly different between images; see text below) has no effect on the travel time, since the wavefront is nearly planar far from the source. In fact, the curvature of the spherical wavefront only perturbs the travel time by
\ba
\frac{(v_\perp\,T_{\rm obs})^2}{2\,c\,D} & \approx & 10^{-3}\,{\rm ms}\,\left( \frac{v_{\perp}}{1000\,{\rm km\,s}^{-1}} \right)^2\,\left( \frac{T_{\rm obs}}{5\,{\rm yr}} \right)^2\,\left( \frac{1\,{\rm Gpc}}{D} \right),
\ea
which is entirely negligible even for transverse velocity
$v_\perp$ as large as the typical cosmic peculiar
velocity. However, lensing deflection causes individual images to
deviate from the unlensed source direction. This offset generates
additional arrival-time perturbation from the Earth's transverse
motion\footnote{That is to say, the decomposition into line-of-sight
  motion and transverse motion differs slightly from one image to
  another, whose effect on the pulse travel time must be
  accounted for.}. Since the size of the Earth's orbit is much smaller
than the typical transverse length scale on the lens plane $\sim
D\,\Delta\theta \approx 5\,{\rm kpc}$, this perturbs the time delay by
an amount that can be estimated by linear variation. This induces a
sinusoidal perturbation to the time delay, whose amplitude is
\ba
\frac{2\,{\rm AU}\,\Delta \theta}{c} \approx 5\,{\rm ms}\,\left( \frac{\Delta\theta}{1''} \right),
\ea
up to an order-unity factor dependent on the orientation of the line of sight with respect to the Earth's orbit. Interestingly, this is potentially measurable given the short FRB width $\sim {\rm ms}$. However, this effect is degenerate with the effect from the line-of-sight projection of the Earth's orbital motion, \refeq{orbitalLoS}, which is sinusoidal with exactly the same period but has an amplitude $\sim 10^5$ times greater! As we will discuss, measurement of time delay perturbation by the Earth's transverse orbital motion would require precise knowledge of the Earth's orbit to an accuracy better than $\sim 10^{-5}$, as well as image localization to $\lesssim 10^{-5}\,{\rm rad} \approx 2''$. Throughout this work, we assume that the former is the case, while the latter is achievable with VLBI observations.

For a ground-based telescope, the Earth's rotation induces a diurnal variation in the time delay in a similar fashion. While the effect from the transverse velocity component coupled to image separation is much smaller than $1\,{\rm ms}$, the line-of-sight velocity component creates a signature as large as $2\,R_\oplus/c \approx 40\,{\rm ms}$. Realistic data analysis will have to account for the Earth's rotation in a similar way to how the Earth's orbital motion is dealt with, but in this work we will neglect this effect for simplicity.

\subsection{Motions of the source/lens}

The source and the lens galaxy typically have cosmic peculiar velocities $\sim \mathcal{O}(1000)\,{\rm km\,s}^{-1}$  with respect to the Solar System, with velocity components both along and perpendicular to the line of sight. Over the observational time span, those can be treated as constant velocities. 

Line-of-sight motion of the source and that of the lens change the radial distances of the source-lens-observer configuration. However, the accumulative change over the typical observation timescale $T_{\rm obs}$ is minuscule compared to $D$. Taking the source's motion as an example, a line-of-sight velocity component $V_{s \parallel}$ induces a change in the time delay
\ba
\label{eq:EQ1}
\Delta T\,\frac{V_{s \parallel}\,T_{\rm obs}}{D} \approx  0.02\,{\rm ms}\,\left( \frac{\Delta T}{0.1\,{\rm yr}} \right)\,\left( \frac{V_{s \parallel}}{1000\,{\rm km\,s}^{-1}} \right)\,\left( \frac{T_{\rm obs}}{5\,{\rm yr}} \right)\,\left( \frac{1\,{\rm Gpc}}{D} \right),
\ea
through the dependence of lensing time delay on the line-of-sight
distances. This is negligible compared to FRB burst widths $\sim {\rm
  ms}$. Even if resolvable, this contributes to the linear drift in
$\Delta T$, distinct from the other non-uniform relative motions we
will focus on in this paper. The same conclusion can be drawn for the
line-of-sight motion of the lens galaxy. 

On the other hand, motion of the source and of the lens perpendicular to the line of sight affect the time delay through a change in the lensing impact parameter. Dominated by the cosmic peculiar velocity, the source's motion induces a linear drift in the time delay
\ba
\label{eq:lineardrift}
\frac{T_{\rm obs}\,v_{s\perp}}{D\,\Delta\theta}\,\Delta T \approx 3\,{\rm s}\, \left( \frac{T_{\rm obs}}{5\,{\rm yr}} \right)\,\left( \frac{v_{s\perp}}{1000\,{\rm km\,s}^{-1}} \right)\,\left( \frac{1\,{\rm Gpc}}{D} \right)\,\left( \frac{1''}{\Delta\theta} \right)\,\left( \frac{\Delta T}{0.1\,{\rm yr}} \right),
\ea
and a similar effect for the lens. These will be measurable as they are much larger than the typical burst width. \refeq{lineardrift} is estimated from linear variation; the correction at quadratic order will be further suppressed by a factor $(T_{\rm obs}\,v_{s\perp})/(D\,\Delta\theta) \sim 10^{-6}$, which is negligible.

The linear drift due to constant transverse velocities is guaranteed to exist, but in individual lensing case it offers limited insight because cosmic peculiar velocities are only statistically predictable. By contrast, any non-uniform transverse motion would be of greater interest. 

It is unclear whether the source of an FRB has significantly non-uniform motion within its host. If the source has an orbital motion because of proximity to, e.g. a stellar companion or a massive black hole, a non-linear perturbation to the time delay is possible\footnote{A somewhat related application is to measure the proper motion of pulsars perpendicular to the line of sight from the scintillation rate \citep{lyne1982interstellar}, and furthermore measure the orbital motion of binary pulsars through sinusoidal  modulations in the scintillation rate \citep{lyne1984orbital}.}. If the source's orbital period is comparable or shorter than the observation time $T_{\rm obs}$, one may search for oscillation in the time delay on the order of
\ba
\label{eq:s-orb}
\frac{a_s}{D\,\Delta\theta}\,\Delta T \approx 3\,{\rm ms}\,\left( \frac{a_s}{1\,{\rm AU}} \right)\,\left( \frac{\Delta T}{0.1\,{\rm yr}} \right)\,\left( \frac{1\,{\rm Gpc}}{D} \right)\,\left( \frac{1''}{\Delta\theta} \right),
\ea
where $a_s$ is the semi-major axis of the source's orbit. If the orbital period is much longer than $T_{\rm obs}$, acceleration may still leave a non-trivial imprint in the time delay, distinct from that of a constant motion.

Another possibility is that FRBs are emitted from numerous compact regions within an extended volume of space. In this case, different repetitions might originate from different regions, whose transverse separations translate into differences in the time delay
\ba
\label{eq:s-rand}
\frac{d_s}{D\,\Delta\theta}\,\Delta T \approx 3\,{\rm ms}\,\left( \frac{d_s}{1\,{\rm AU}} \right)\,\left( \frac{\Delta T}{0.1\,{\rm yr}} \right)\,\left( \frac{1\,{\rm Gpc}}{D} \right)\,\left( \frac{1''}{\Delta\theta} \right),
\ea
where $d_s$ is the typical transverse separation between emission
regions. Thus, separations as small as a fraction of $1\,{\rm AU}$ can
be detectable, which by far exceeds the resolution of radio interferometry. Again, \refeq{s-orb} and \refeq{s-rand} are based on linear variation of the lensing impact parameter; quadratic corrections are negligibly small.

The above back-of-the-envelope estimates suggest that repeating FRBs, if strongly lensed into multiple images, may provide us with unique opportunities to measure time delay perturbations induced by the motions of the source and the Earth, thanks to their narrow burst widths. With information on the source's motion, much may be learned about its physical properties and its surrounding environment. In the following, we derive rigorous equations.

\section{Time delay perturbations}
\label{sec:timedelay}

\begin{figure}[t]
  \begin{center}
   \hspace{-0.7cm}
    \includegraphics[scale=0.55]{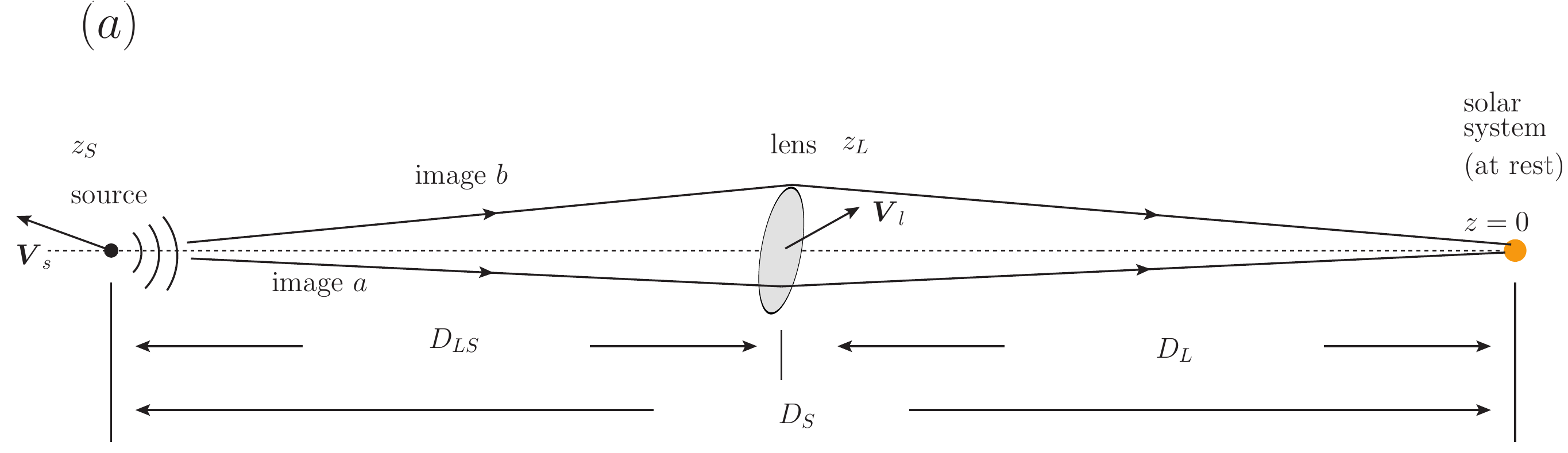} \\
    \vspace{0.5cm}
    \includegraphics[scale=0.55]{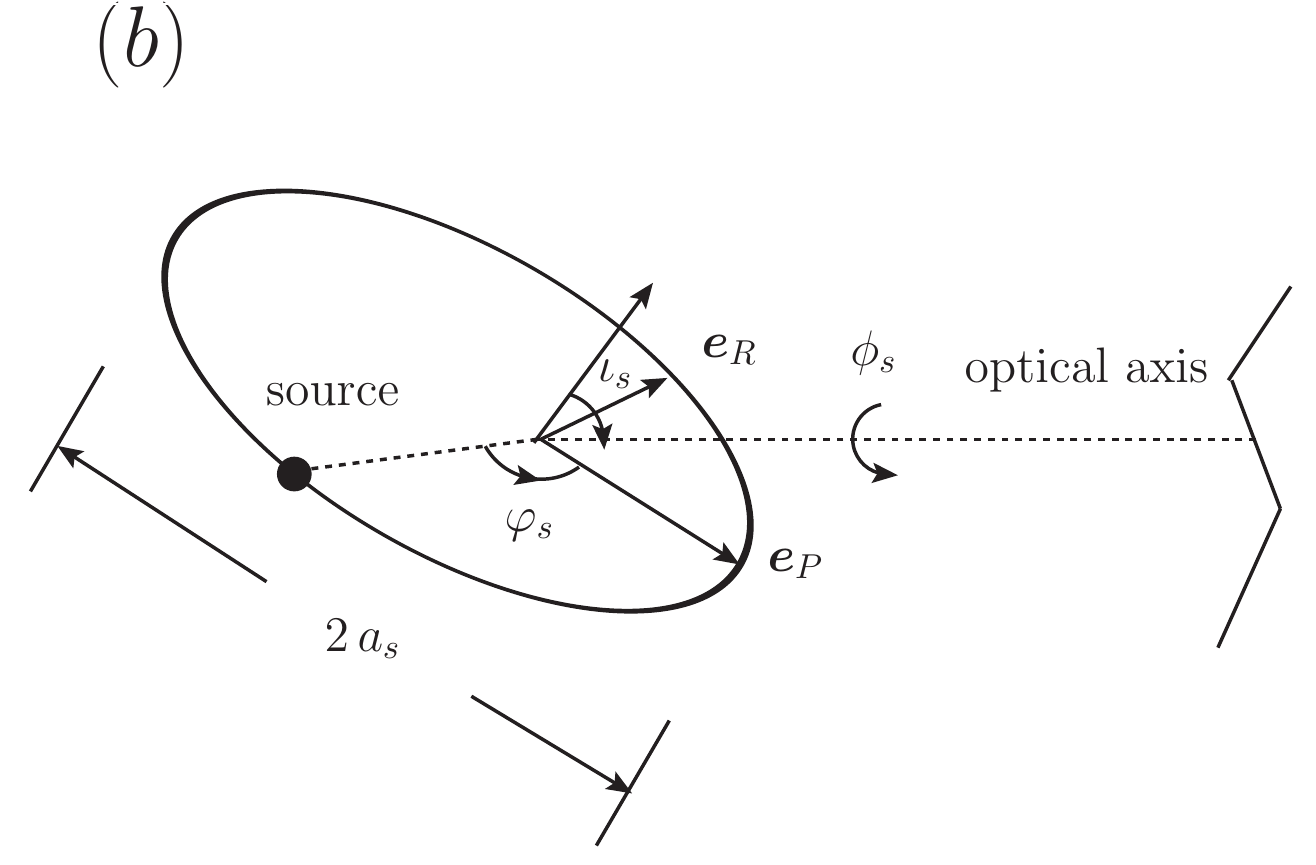}\quad
    \includegraphics[scale=0.55]{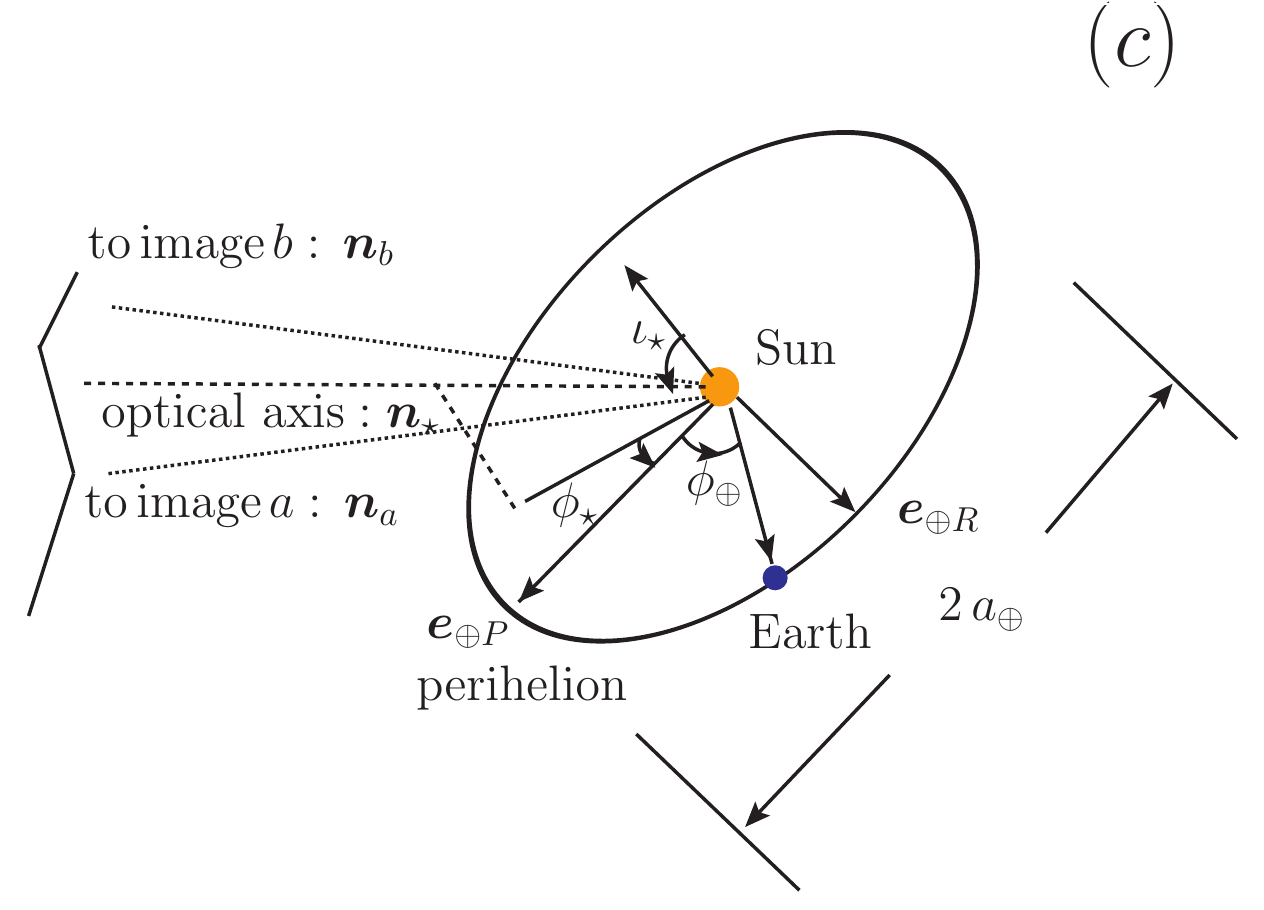}
    \caption{\label{fig:lensing_geometry} Geometrical configuration of lensing discussed in this work. (a) Alignment of the FRB source, the lens galaxy, and the solar system along the optical axis. The source and the lens have quasi-uniform peculiar motions with respect to the Solar System. (b) The possible scenario of a source in a Keplerian orbit. (c) The orbital motion of the Earth around the Sun in the rest frame of the Solar System. The angular difference between different lensing images is exaggerated. }
  \end{center}
\end{figure}

In this section, we rigorously derive the aforementioned kinematic effects on the lensing time delay. We first assume an observer at rest with respect to the barycenter of the Solar System. In the end we discuss how to convert observables into the Earth's rest frame.

Consider the strong lensing geometry as in
\reffig{lensing_geometry}(a): a repeating FRB is located at redshift $z_S$ with an angular-diameter distance $D_S$ to the Earth. An intervening galaxy lens is located at redshift $z_L < z_S$ with an angular-diameter distance $D_L$ to the Earth. The angular-diameter distance to the source as viewed from the lens is $D_{LS}$.

Let $\bfy$ be the true (angular) position of the source, and $\bfx$ be the image (angular) position on the lens plane, and $\bfz$ be the (angular) position of the lens galaxy. These are measured relative to a reference line of sight, which we call the {\it optical axis}. We first allow $\bfz \neq 0$ for the convenience of considering lens motion; after calculations are done, we are free to set $\bfz=0$. For a stationary lensing configuration, the Fermat potential is given by~\citep{schneider1992gravitational}
\ba
F(\bfx;\,\bfy,\bfz) & = & \left( 1+z_L \right)\,\frac{D_L\,D_S}{c\,D_{LS}}\,\left[ \frac12\,\left( \bfx - \bfy \right)^2  - \psi\left(\bfx - \bfz \right)  \right],
\ea
where $\psi(\bfx)$ is the usual lensing potential if the lens's center is right on the optical axis. This is defined relative to the geometrical travel time from the source to the Earth along a direct straight line.

Images are located at the extremal points of the Fermat potential, which we label by $I=a, b, c, \cdots$. Their positions $\bfx_I$ are roots of the lens equation,
\ba
\label{eq:lenseq}
\bfx - \bfy - \bfalp\left(\bfx - \bfz \right) = 0.
\ea
Here, the deflection angle $\bfalp(\bfx-\bfz)$ equals the gradient of the lensing potential $\bfalp(\bfx-\bfz) = \partial \psi(\bfx-\bfz)/\partial \bfx$, where $\psi(\bfx-\bfz)$ is linearly proportional to the Shapiro time delay due to the gravitational field of the lens. The observed time delay of the $J^{\rm th}$ image relative to the $I^{\rm th}$ image is given by
\ba
\label{eq:Tij0}
\Delta T_{IJ,0} & = & \Delta T_{J,0} - \Delta T_{I,0}, \en
\Delta T_{I,0} & = & \left( 1+z_L \right)\,\frac{D_L\,D_S}{c\,D_{LS}}\,\left[ \frac12\,\left( \bfx_I - \bfy \right)^2  - \psi\left( \bfx_I - \bfz \right) \right],
\ea
and similarly for $\Delta T_{J,0}$. We have introduced a subscript ${}_0$ to remind ourselves that this assumes no relative motion between the source, the lens, and the Earth. 

Assume that the source gives off successive bursts, labeled by $k=1,2,3,\cdots$. For each burst, the same set of multiple images is generated.

\subsection{Earth's orbital motion}

First, we study the effect of the Earth's orbital motion around the Sun. As in \reffig{lensing_geometry}(c), suppose that relative to the normal of the Earth's orbital plane, the true position of the source on the sky (as seen in the inertial frame of the barycenter of the Solar System) has an inclination angle $\iota$, and it has an azimuthal angle $\phi$ relative to the direction of perihelion. In the same coordinate system, the optical axis has inclination $\iota_\star$ and azimuthal angle $\phi_\star$. Note that $(\iota, \phi)$ and $(\iota_\star, \phi_\star)$ are {\it not} identical. They differ by a small angular displacement $\bfy$.

Since the radio wave coming from each image has a (nearly) planar wavefront when reaching the Solar System, the correction to time delay, for the $k^{\rm th}$ burst, is given by
\ba
\label{eq:delTkijEarth}
\delta T^{(k)}_{I,\,\oplus} = - \frac{1}{c}\, \bfn_I \cdot \bfr_\oplus(t^{(k)}_I), \qquad \delta T^{(k)}_{IJ,\,\oplus} = \delta T^{(k)}_{J,\,\oplus}(t^{(k)}_J) - \delta T^{(k)}_{I,\,\oplus}(t^{(k)}_I).
\ea
Here $\bfn_I$ is the unit vector to the $I^{\rm th}$ image on the sky. And we have used $t^{(k)}_I$ to denote the time of arrival for the $I^{\rm th}$ image of the $k^{\rm th}$ burst.

The vector $\bfr_\oplus(t)$ is the Earth's displacement in three-dimensional space at given time $t$. It can be written as
\ba
\label{eq:rkij}
\bfr_\oplus(t) & = & \frac{p_\oplus}{1 + e_\oplus\,\cos\phi_\oplus(t)}\,\left[ \bfe_{\oplus P}\,\cos\phi_\oplus(t) + \bfe_{\oplus R}\,\sin\phi_\oplus(t)  \right].
\ea 
Here $e_\oplus = 0.0167$ is the orbital eccentricity, $a_\oplus =
1.496 \times 10^8\,{\rm km}$ is the semi-major axis, $p_\oplus =
a_\oplus (1-e^2_\oplus)$ is the semi-latus rectum, and
$\phi_\oplus(t)$ gives the azimuthal position of the Earth relative to
the perihelion at a given time $t$. We also define $\bfe_{\oplus P}$ to be a unit vector pointing from the Sun to the perihelion, and $\bfe_{\oplus R}$ is another unit vector in the orbital plane orthogonal to $\bfe_{\oplus P}$. Since the orbital eccentricity is small, $\phi_\oplus(t) \approx \Omega_\oplus\,t$, where $\Omega_\oplus$ is the angular frequency of the Earth's orbital motion.

It is convenient to decompose $\bfr_\oplus(t)$ into a component parallel to the optical axis $\bfd_\oplus(t)$ and a component perpendicular to it,
\ba
\bfr_\oplus(t) = \bfd_\oplus(t) + l_\oplus(t)\,\bfn_\star, \qquad l_\oplus(t) = \bfr_\oplus(t)\cdot\bfn_\star, \qquad \bfd_\oplus(t) = \bfr_\oplus(t) - \left[ \bfr_\oplus(t)\cdot\bfn_\star \right]\,\bfn_\star,
\ea
using a unit vector $\bfn_\star$ pointing along the optical axis\footnote{The decomposition of a three-dimensional vector into parallel and transverse components artificially depends on the choice of a reference ``line-of-sight'' direction. Here $\bfn_\star$ pointing along the optical axis is chosen. Once such a choice is made, following calculations should be done consistently.}. Since $\bfn_I = \bfn_\star + \bfx_I$, \refeq{delTkijEarth} can be decomposed into an effect due to line-of-sight motion, and an effect due to transverse motion,
\ba
\delta T^{(k)}_{I,\oplus} & = & \delta T^{(k)}_{I,\oplus\parallel} + \delta T^{(k)}_{I,\oplus\perp}, \en
\delta T^{(k)}_{I,\oplus\parallel} & = & - \frac{1}{c}\,l_\oplus(t^{(k)}_I), \en
\delta T^{(k)}_{I,\oplus\perp} & = & - \frac{1}{c}\,\bfx_I\cdot\bfd_\oplus(t^{(k)}_I),
\ea
where we have ignored terms quadratic in the small angles $\bfx_I$'s (which generate minuscule fractional corrections $\sim (\Delta\theta)^2 \sim 10^{-12}$). The effect due to line-of-sight motion therefore reads
\ba
\label{eq:delTkijEarthpara}
\delta T^{(k)}_{IJ,\,\oplus \parallel} = \delta T^{(k)}_{J,\,\oplus \parallel} - \delta T^{(k)}_{I,\,\oplus \parallel} = \frac{p_\oplus\,\sin\iota_\star}{c}\, \left[ \frac{\cos\left(\phi_\star - \phi_\oplus(t^{(k)}_J)\right)}{1+e_\oplus\,\cos\phi_\oplus(t^{(k)}_J)} - \frac{\cos\left(\phi_\star - \phi_\oplus(t^{(k)}_I)\right)}{1+e_\oplus\,\cos\phi_\oplus(t^{(k)}_I)} \right].
\ea
On the other hand, the effect due to transverse motion is given by
\ba
\label{eq:delTkijperp}
\delta T^{(k)}_{IJ,\,\oplus\perp} = \delta T^{(k)}_{J,\,\oplus\perp} - \delta T^{(k)}_{I,\,\oplus\perp} = \frac{1}{c} \left[ \bfx_I \cdot \bfd_{\oplus}(t^{(k)}_I) - \bfx_J \cdot \bfd_{\oplus}(t^{(k)}_J) \right],
\ea
which only depends on image positions $\bfx_I$ but not on details of the lens.

\subsection{Motion of the lens}

As in \reffig{lensing_geometry}(a), the lens galaxy has a constant peculiar velocity $\bfV_l$ relative to the Solar System. According to \refsec{order}, only the velocity component transverse to the optical axis induces a sizable perturbation to the time delay. To derive this effect, consider differentiation of $F(\bfx; \bfy, \bfz)$ with respect to $\bfz$,
\ba
\frac{d \Delta T_I}{d t_l} = \frac{V_{l\perp}^i}{D_L} \, \frac{\partial F\left(\bfx_I;\,\bfy,\bfz\right)}{\partial z^i},
\ea
where $t_l$ is the {\it lens-frame} time (measured relative to a chosen reference moment $t_l=0$) satisfying $dt = (1+z_L)\,dt_l$ due to cosmic time dilation. When computing the derivative, we fix $\bfy$ but $\bfx$ is regarded dependent on $\bfz$ via the lens equation \refeq{lenseq},
\ba
\frac{\partial}{\partial z^i}\,\psi\left(\bfx-\bfz\right) = \alpha_j\left( \bfx-\bfz \right)\,\left( \frac{\partial x^j}{\partial z^i} - \delta^j{}_i \right).
\ea
Combining these results with the lens equation, integrating over $t_l$, we obtain (and set $\bfz=0$ eventually)
\ba
\label{eq:delTkIlperp}
\delta T^{(k)}_{I,l\perp} & = & (1+z_L)\, \frac{D_S}{c\,D_{LS}}\,\bfalp_I\cdot \bfd_l(t^{(k)}_{l J}), \\
\delta T^{(k)}_{IJ,l\perp} & = & \delta T^{(k)}_{J,l\perp} - \delta T^{(k)}_{I,l\perp} = (1+z_L)\, \frac{D_S}{c\,D_{LS}}\,\left[ \bfalp_J \cdot \bfd_l(t^{(k)}_{l J}) - \bfalp_I \cdot \bfd_l(t^{(k)}_{l I}) \right].
\ea
Here $\bfd_l(t_l)$ is the linear displacement of the lens transverse to the line of sight at given lens-frame time $t_l$. The deflection $\bfalp_I = \bfx_I - \bfy$ is not directly measurable without knowing the true source position $\bfy$. The lens-frame time $t^{(k)}_{lI}$ when the $I^{\rm th}$ image of the $k^{\rm th}$ radio burst passes the lens may be chosen to be
\ba
\label{eq:lens-time}
t^{(k)}_{lI} = \frac{t^{(k)}_0}{1 + z_L} + \frac{(1+z_L)}{2\,(1+z_S)\,c}\,\frac{D^2_L}{D_{LS}}\,\left(\bfx_I - \bfy \right)^2 - \frac{D_L\,D_S}{2\,c\,D_{LS}}\,\psi(\bfx_I).
\ea
The definition of $t^{(k)}_{lI}$ is arbitrary within the light-crossing time of the lens galaxy. The bottom line, however, is that the effect of the lens's motion is to high precision merely a linear drift in the burst time of arrival in the observer's frame, if we assume that $\bfV_l$ is constant.

\subsection{Motion of the source}

The source also has a velocity $\bfV_s$ relative to the Solar System. As explained in \refsec{order}, only the velocity component transverse to the optical axis is relevant. For the source, we first set $\bfz=0$, and then compute the linear variation of $F(\bfx;\,\bfy,\bfz=0)$ with respect to $\bfy$,
\ba
\frac{d \Delta T_s}{d t_s} =\frac{V_{s\perp}^i}{D_S} \, \frac{\partial F\left(\bfx_I;\,\bfy \right)}{\partial y^i},
\ea
where $t_s$ is the {\it source-frame} time (measured relative to a chosen reference moment $t_s=0$) satisfying $dt = (1+z_S)\,dt_s$. When computing the derivative, we treat $\bfx$ as dependent on $\bfy$ via the lens equation \refeq{lenseq}, and find
\ba
\frac{\partial\,\psi(\bfx)}{\partial y^i} = \frac{\partial\,\psi(\bfx)}{\partial\,x^j}\,\frac{\partial\,x^j}{\partial\,y^i} = \alpha_j(\bfx)\,\frac{\partial\,x^j}{\partial\,y^i}.
\ea
Combining these results with the lens equation, integrating over $t_s$, we derive the perturbation in time delay~\citep{1999ApJ...519L..31Y}
\ba
\label{eq:delTkijs}
\delta T^{(k)}_{I,s\perp} & = & -  (1+z_L)\,\frac{D_L}{c\,D_{LS}}\,\bfalp_I \cdot \bfd_s(t^{(k)}_{s}), \\
\label{eq:delTkijs_mutual}
\delta T^{(k)}_{IJ,s\perp} & = & \delta T^{(k)}_{J,s\perp} - \delta T^{(k)}_{I,s\perp} = - (1+z_L)\,\frac{D_L}{c\,D_{LS}}\,\left( \bfx_J  - \bfx_I \right) \cdot \bfd_s(t^{(k)}_{s}).
\ea
where $\bfd_s(t_s)$ is the transverse displacement of the source at a given source-frame time $t_s$, and $t^{(k)}_{s}$ is the moment in the source frame when the $k^{\rm th}$ burst is emitted, which is the same for all images. The vector $\bfd_s(t_s)$ can be obtained by projecting the three-dimensional displacement of the source $\bfr_s(t_s)$ onto the plane perpendicular to the line of sight, namely $\bfd_s(t_s) = \bfr_s(t_s) - [\bfr_s(t_s) \cdot \bfn_\star]\,\bfn_\star $. Note that this result depends on $\bfx_I$'s, which are direct observables, but not on the lens model.

The displacement of the source $\bfr_s(t_s)$ is expected to be dominated by a constant cosmic peculiar velocity $\bfV_s$ relative to the lens. If in addition the source has Keplerian motion orbiting a massive object (\reffig{lensing_geometry}(b)), described by an orbital eccentricity $e_s$, the semi-latus rectum of the orbit $p_s$, which is related to the semi-major axis $a_s$ through $p_s = a_s\,(1-e_s^2)$, and the instantaneous azimuthal position in the orbital plane $\phi_s(t_s)$ (for simplicity, we neglect possible orbital precession), we can then write
\ba
\label{eq:rsts}
\bfr_s(t_s) = \bfV_s\,t_s + \frac{p_s}{1 + e_s\,\cos\phi_s(t_s)}\,\left[ \bfe_P\,\cos\phi_s(t_s) + \bfe_R\,\sin\phi_s(t_s)  \right].
\ea 
Here $\bfe_P$ is a unit vector pointing from the binary center of mass to the periapsis, and $\bfe_R$ is another unit vector in the orbital plane orthogonal to $\bfe_P$. Analogous to the case of the Earth's motion, after \refeq{rsts} is inserted into \refeq{delTkijs}, the first term generates a linear drift in the time delay, while the second term induces an oscillatory perturbation.

In summary, the mutual time delay between a given pair of images belonging to the $k^{\rm th}$ burst is given by the zeroth-order delay $\Delta T_{IJ,0}$ computed for a stationary lensing configuration, further corrected by various velocity effects,
\ba
\label{eq:totaldelay}
\Delta T^{(k)}_{IJ} & = & \Delta T_{IJ,0} + \delta T^{(k)}_{IJ,\oplus\parallel} + \delta T^{(k)}_{IJ,\oplus\perp} + \delta T^{(k)}_{IJ,l\perp} + \delta T^{(k)}_{IJ,s\perp}.
\ea
Among them, the effect of the Earth's motion results from the finite light travel time across the Earth's orbit. By comparison, the effects of transverse motions for the lens and for the source depend on image separations and can be understood on the basis of a change in the lensing impact parameter.

\subsection{Observing in the Earth's rest frame}

We have performed calculations in the rest frame of the Solar System barycenter, while radio telescopes co-move with the Earth, which defines a non-inertial frame. Moreover, a few relativistic effects may need to be accounted for if high precision is desired.

Bursts are timed by a clock co-moving with the Earth, which is slightly slower than a clock in the inertial frame of the Solar System, due to both kinetic and gravitational time dilation. However, dilation rescales all time intervals in the same way, so that it produces no drift or oscillation.

Localization using telescopes on the Earth are subject to relativistic aberration. The apparent position of an image annually traces an ellipse on the sky, whose semi-major axis is $\sim 20''$ and whose semi-minor axis depends on the Ecliptic latitude. This affects the image coordinates for different repetitions at different times of the year. Aberration may be unimportant if sky localization is poor (and hence far from sufficient to resolve images), but needs to be corrected for with VLBI resolution. Annual aberration modulates the absolute, apparent position in the Ecliptic coordinates, but (nearly) preserves the angular separations between the images, the source, and the lens. In the following, we will assume that aberration due to the Earth's orbital motion is always corrected for. The same can be done for diurnal aberration due to the Earth's rotation.

\section{Simulating mock data}
\label{sec:simulation}

To demonstrate how well the source's motion can be inferred from the aforementioned effects on the time delay, we construct a hypothetical strong-lensing event and simulate mock measurements. This example will be adopted throughout.

We hypothesize a repeater at $z_S = 1.0$, which is strongly lensed by an intervening galaxy at $z_L=0.5$. Assuming the Planck best-fit cosmological parameters \citep{Ade:2015xua}, we obtain angular diameter distances $D_S = 1635\,{\rm Mpc}$, $D_L = 1251\,{\rm Mpc}$, and $D_{LS} = 697\,{\rm Mpc}$. As a concrete example, the lens galaxy is assumed to be a singular isothermal ellipsoid (SIE)~\citep{1994A&A...284..285K} with a velocity dispersion $\sigma_v = 250\,{\rm km\,s}^{-1}$ and an axis ratio $f=0.4$. For simplicity, we ignore the possibility of an external shear.

Without lensing deflection, we assume that the line of sight to the source has an inclination $\iota = 0.5$ with respect to the normal of the Ecliptic plane, and an azimuthal position $\phi = 0.7$ relative to the Earth's perihelion.

\begin{figure}[t]
  \begin{center}
   \hspace{-0.7cm}
    \includegraphics[scale=0.65]{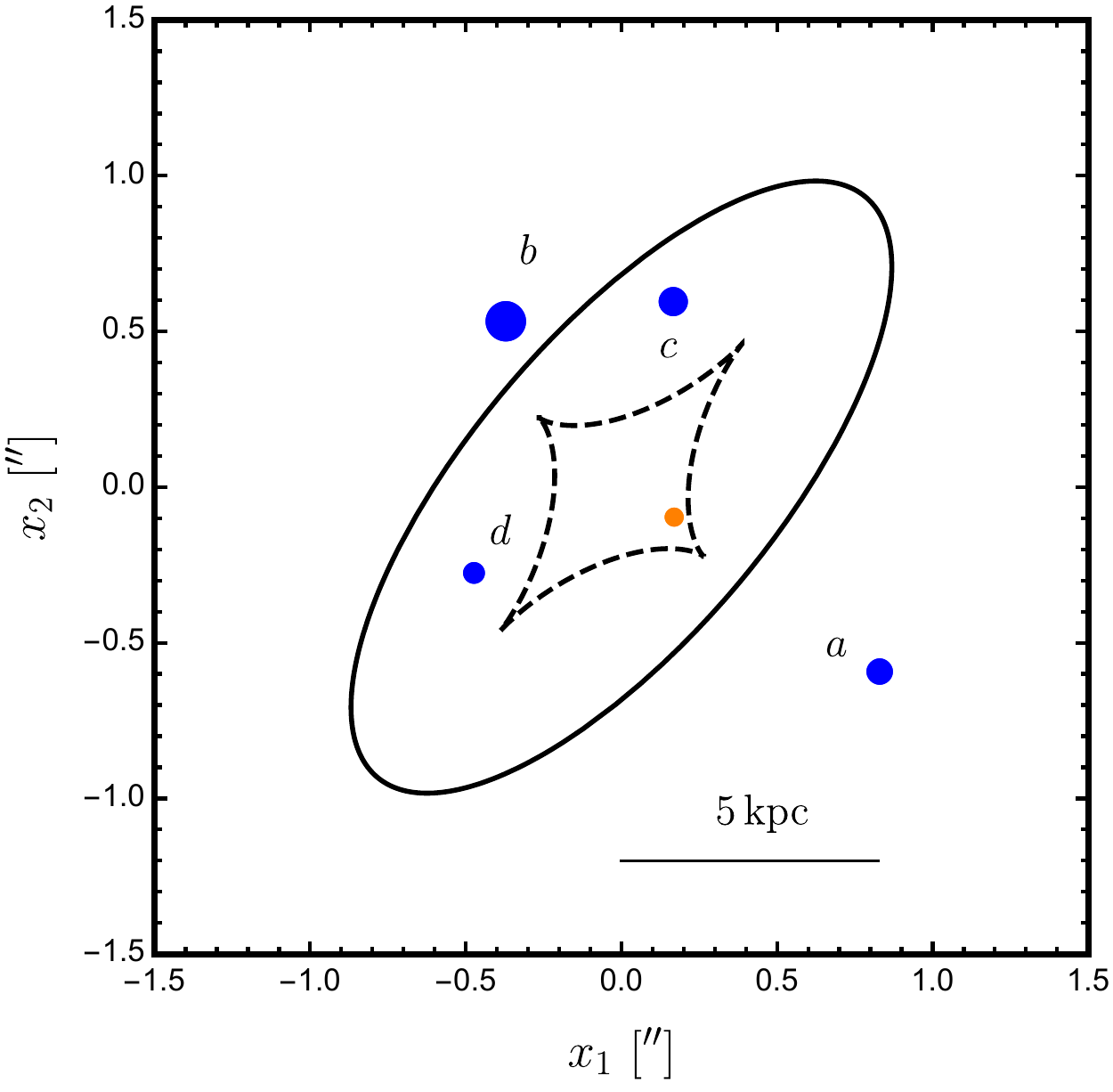}
    \caption{\label{fig:lens} Location of the hypothetical source (orange disk) and of its four images (blue disks). Circles have their areas proportional to the geometrical magnification factor. Critical curve (solid) and caustic (dashed) are also shown. The coordinate origin is chosen to be the geometrical center of the hypothetical lens. A linear scale on the lens plane is also shown for reference.}
  \end{center}
\end{figure}

On the sky, we set up a Cartesian coordinate system centered at the geometrical center of the lens, whose first coordinate axis is parallel to the Ecliptic plane. Its major axis on the sky makes an angle $\varphi_L = 0.7$ relative to the direction parallel to the Ecliptic plane. Under this coordinate system, we assume a source at angular location $\bfy=( 0.223, \,-0.123 )$, in units of the characteristic angular scale $\xi_0 = 4\,\pi\,(\sigma_v/c)^2\,D_{LS}/D_S = 0.770''$. If the source-lens-observer configuration is stationary, four images are produced in geometrical optics (\reffig{lens}), whose angular locations and mutual time delays are presented in \reftab{image_list}. Among them, $a$ precedes $b$, $c$ and $d$ by about $40$ days, while the latter three images have mutual time delays on the order of a few days.

We assume that the source has a constant velocity relative to the Solar System, with a component transverse to the optical axis $\bfV_{s\perp} = (\cos\vartheta_s, \sin\vartheta_s)\times 1200\,{\rm km\,s}^{-1}$ with $\vartheta_s = 0.8$, and that the lens galaxy also moves at a constant velocity relative to the Solar System, with a component transverse to the optical axis, $\bfV_{l \perp} = (\cos\vartheta_l, \sin\vartheta_l)\times 800\,{\rm km\,s}^{-1}$ with $\vartheta_l=2.5$. As has been explained in \refsec{order}, the line-of-sight components are not relevant. 

\begin{table}[]
\begin{center}
\setlength\tabcolsep{9pt}
\begin{tabular}{l|c|c|c|c|cl|}
\specialrule{.1em}{.05em}{.05em} 
 Image & $\bfx_I\,['']$ & $\bfalp(\bfx_I)\,['']$ & $\Delta T_{aI,0}\,[{\rm day}]$ & $\mu_I$  \\
\hline
\hline
$a$ & ( 0.831,\,-0.604 ) & (0.659,\,-0.509) & --- & 1.91 \\
$b$ & ( -0.373,\,0.526 ) & (-0.546,\, 0.621) & 39.392 & 4.46 \\
$c$ & ( 0.167,\,0.596 ) & (-0.005,\,0.690) & 41.453 & 2.33 \\
$d$ & ( -0.472,\,-0.276 ) & (-0.644,\,-0.182) & 46.601 & 1.31 \\
\specialrule{.1em}{.05em}{.05em} 
\end{tabular}
\caption{\label{tab:image_list} Angular positions $\bfx_I$ of the image quad, their time delays relative to the first image, and the associated magnification factors $\mu_I$.}
\end{center}
\end{table}

Timing of bursts is then simulated according to the following procedure:

\begin{enumerate}

\item For simplicity, a series of repetition times $t^{(k)}_0$ for
  $k=1,2,\cdots$ are generated according to a random Poisson process
  with a constant repetition rate (albeit in reality the repeater
  FRB121102~\citep{Spitler:2016dmz} shows remarkable non-Poissonian
  behavior~\citep{Wang:2016lhy,Opperman:2017kql}). These $t^{(k)}_0$'s serve as reference times of arrival in observer's frame for individual repetitions, corresponding to the case {\it if strong lensing did not happen and if the source, the lens, and the Earth were not moving}, i.e. radio waves travel directly along straight lines.

\item 
Forcing emission at the source and reception at the Earth in general leads to an implicit equation, which we solve in the following iterative way to obtain sufficiently accurate arrival times. For the $I^{\rm th}$ image of the $k^{\rm th}$ burst, a {\it trial} time of arrival is first computed according to
\ba
\label{eq:toa_trial}
\tilde t^{(k)}_I & = & t^{(k)}_0 + \Delta T_{I,0} + \frac{p_\oplus\,\sin\iota_\star}{c}\,\frac{\cos\left(\phi_\star - \phi_\oplus\left(t^{(k)}_0 + \Delta T_{I,0}\right)\right)}{1+e_\oplus\,\cos\phi_\oplus\left(t^{(k)}_0 + \Delta T_{I,0}\right)} - \frac1c\,\bfx_I \cdot \bfd_{\oplus}\left( t^{(k)}_0 + \Delta T_{I,0} \right) \en
&& - (1+z_L)\,\frac{D_L}{c\,D_{LS}}\,\bfx_I \cdot \bfd_s \left( t^{(k)}_{s} \right) + (1+z_L)\,\frac{D_S}{c\,D_{LS}}\,\bfalp_I\cdot\bfd_l\left( t^{(k)}_{lI} \right).
\ea 
Here $t^{(k)}_{s} = t^{(k)}_0/(1+z_S)$ is the time of emission of the $k^{\rm th}$ burst in the source frame, and 
$t^{(k)}_{lI}$ is given by \refeq{lens-time}. We then find $t^{(k)}_I$ by {\it re-calculating} the right hand side of \refeq{toa_trial} and replacing the combination $t^{(k)}_0 + \Delta T_{I,0}$ with the trial $\tilde t^{(k)}_I$ whenever the Earth's instantaneous position needs to be computed, i.e.
\ba
\label{eq:toa}
t^{(k)}_I & = & t^{(k)}_0 + \Delta T_{I,0} + \frac{p_\oplus\,\sin\iota_\star}{c}\,\frac{\cos\left(\phi_\star - \phi_\oplus\left( \tilde t^{(k)}_I \right)\right)}{1+e_\oplus\,\cos\phi_\oplus\left( \tilde t^{(k)}_I \right)} - \frac1c\,\bfx_I \cdot \bfd_{\oplus}\left( \tilde t^{(k)}_I \right) \en
&& - (1+z_L)\,\frac{D_L}{c\,D_{LS}}\,\bfx_I \cdot \bfd_s \left( t^{(k)}_{s} \right) + (1+z_L)\,\frac{D_S}{c\,D_{LS}}\,\bfalp_I\cdot\bfd_l\left( t^{(k)}_{lI} \right).\ea 
In this way, we perturbatively ensure that the answer for $t^{(k)}_I$ is consistent with propagation along a null ray and is accurate to a level better than $\sim 1\,{\rm ms}$.

\item The time delay of the $J^{\rm th}$ image relative to the $I^{\rm th}$ image for the $k^{\rm th}$ burst is readily found by $\Delta T^{(k)}_{IJ} = t^{(k)}_J - t^{(k)}_I$. 

\item Finally, timing noise drawn from a normal distribution with zero mean and a standard deviation $\sigma_w$ is added to the mock times of arrival. This is to account for the limitation of finite burst width on the (de-dispersed) timing accuracy.
 
\end{enumerate}

\section{Localization without resolving images}
\label{sec:localize}

Single-dish telescopes or short-baseline arrays used for radio transient surveys are typically incapable of resolving lensed multiple images. Large interferometric arrays or VLBI technique are therefore needed to pin down the host-lens system and separate images on sub-arcsecond angular scales. However, the coarse localization of survey telescopes is adverse for efficient VLBI follow-ups. We now discuss, in the case of a multiply-imaged repeater, how information on the time-delay perturbation might help improve localization and facilitate deep follow-ups.

As estimated in \refsec{order}, line-of-sight projection of the Earth's orbital motion induces the largest non-trivial perturbation to the time delay $\delta T^{(k)}_{IJ,\oplus\parallel} \lesssim 10^3\,$s. Since $e_\oplus \ll 1$, this has a nearly sinusoidal temporal variation. Assume that the Earth's orbital motion is known to high accuracy, according to \refeq{delTkijEarthpara}, the inclination $\iota$ and azimuthal angle $\phi$ of the source's sky position in the Ecliptic coordinates can be deduced from the amplitude and the phase of this variation.

We now study how precisely one can localize the source in this way. Suppose a total number of $N$ bursts are detected, each of which has multiple images. Since we seek variation on the order of hundreds of seconds, we may neglect non-uniform transverse velocities in $\delta T^{(k)}_{IJ,\oplus\perp}$ and $\delta T^{(k)}_{IJ,s\perp}$. For the $I^{\rm th}$ image and the $J^{\rm th}$ image, we use the following model for the mutual delay
\ba
\label{eq:5par}
\Delta T^{(k)}_{IJ} \equiv t^{(k)}_J - t^{(k)}_I = \Delta T_{IJ,0} + \left( K_J\,t^{(k)}_J - K_I\,t^{(k)}_I \right) + \frac{p_\oplus\,\sin\iota}{c}\, \left[ \frac{\cos\left(\phi - \phi_\oplus(t^{(k)}_J)\right)}{1+e_\oplus\,\cos\phi_\oplus(t^{(k)}_J)} -\frac{\cos\left(\phi - \phi_\oplus(t^{(k)}_I)\right)}{1+e_\oplus\,\cos\phi_\oplus(t^{(k)}_I)} \right],
\ea
which has five free parameters $(\iota,\, \phi,\,\Delta T_{IJ,0},\,K_I,\,K_J)$. Among them, we mainly aim to measure the sky localization $\iota$ and $\phi$ averaged over all images; \refeq{5par} does not account for image separations, since this method cannot achieve sufficient angular resolution to resolve individual images anyway. The other three are nuisance parameters: $\Delta T_{IJ,0}$ describes a constant time delay due to stationary lensing; $K_I$ and $K_J$ account for linear drifts in the time delay induced by constant (but unknown) velocities. Assuming that one has perfect knowledge of the Earth's orbital parameters, and that timing of all bursts have a gaussian random uncertainty $\sigma_w$ due to finite burst widths, we can find the best-fit parameters by maximizing the log likelihood,
\ba
\label{eq:5parlnL}
\ln \mathcal{L} = - \frac{1}{4\,\sigma^2_w}\,\sum^N_{k=1}\, && \Bigg\{ t^{(k)}_J - t^{(k)}_I - \Delta T_{IJ,0} - \,\left( K_J\,t^{(k)}_J - K_I\,t^{(k)}_I \right) \en
&&  - \frac{p_\oplus\,\sin\iota}{c}\, \Bigg[ \frac{\cos\left(\phi - \phi_\oplus(t^{(k)}_J)\right)}{1+e_\oplus\,\cos\phi_\oplus(t^{(k)}_J)} -\frac{\cos\left(\phi - \phi_\oplus(t^{(k)}_I)\right)}{1+e_\oplus\,\cos\phi_\oplus(t^{(k)}_I)} \Bigg]  \Bigg\}^2.
\ea 
The factor of four in front of $\sigma^2_w$ comes from the fact that the difference between two independent timing measurements has a variance $2\,\sigma^2_w$. Similar factors arise in equations presented later. It is worthy to note that this method does not require redshift information of the lens or the source.

\begin{figure}[t]
  \begin{center}
    \includegraphics[scale=0.65]{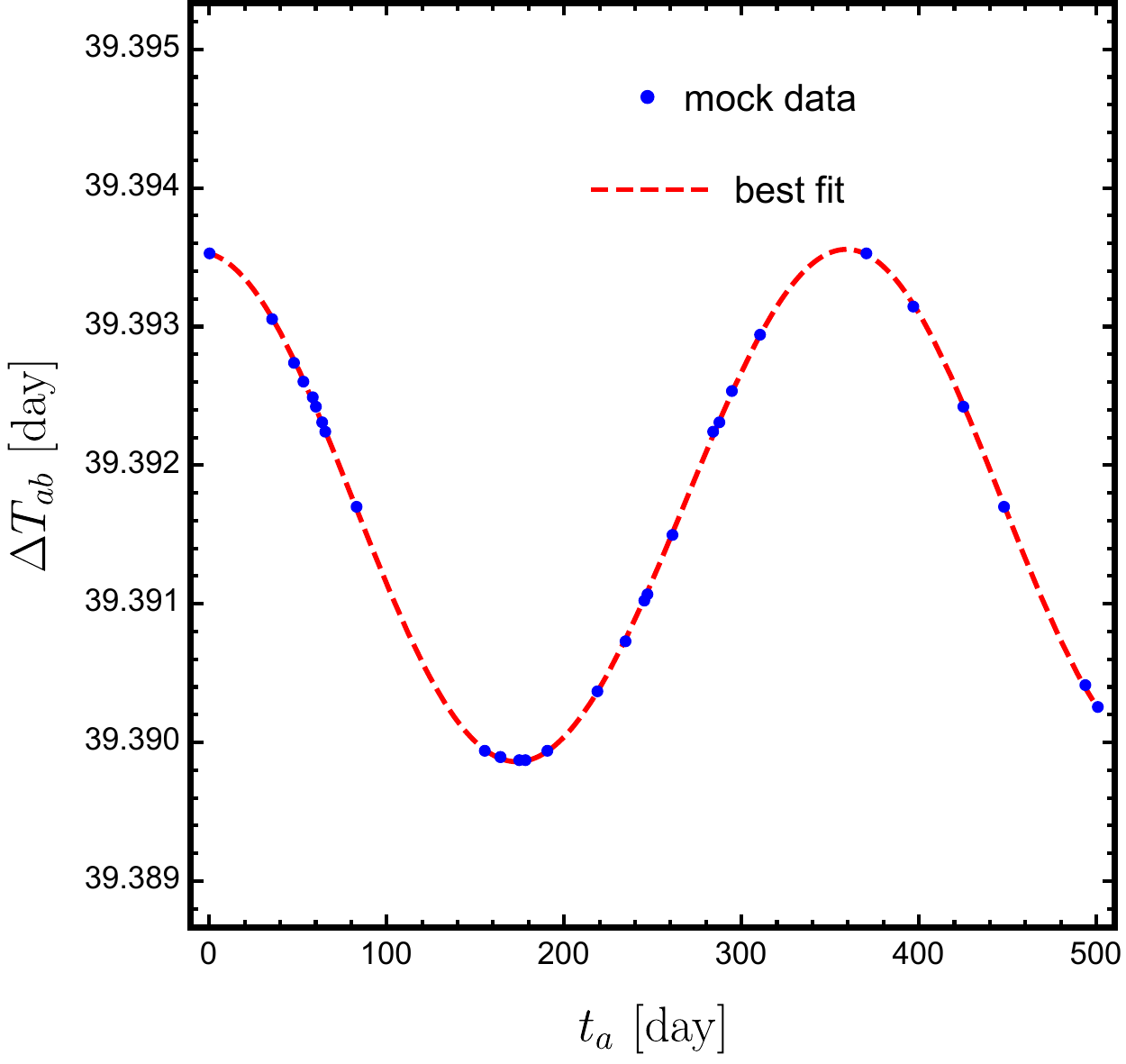}\qquad
    \includegraphics[scale=0.62]{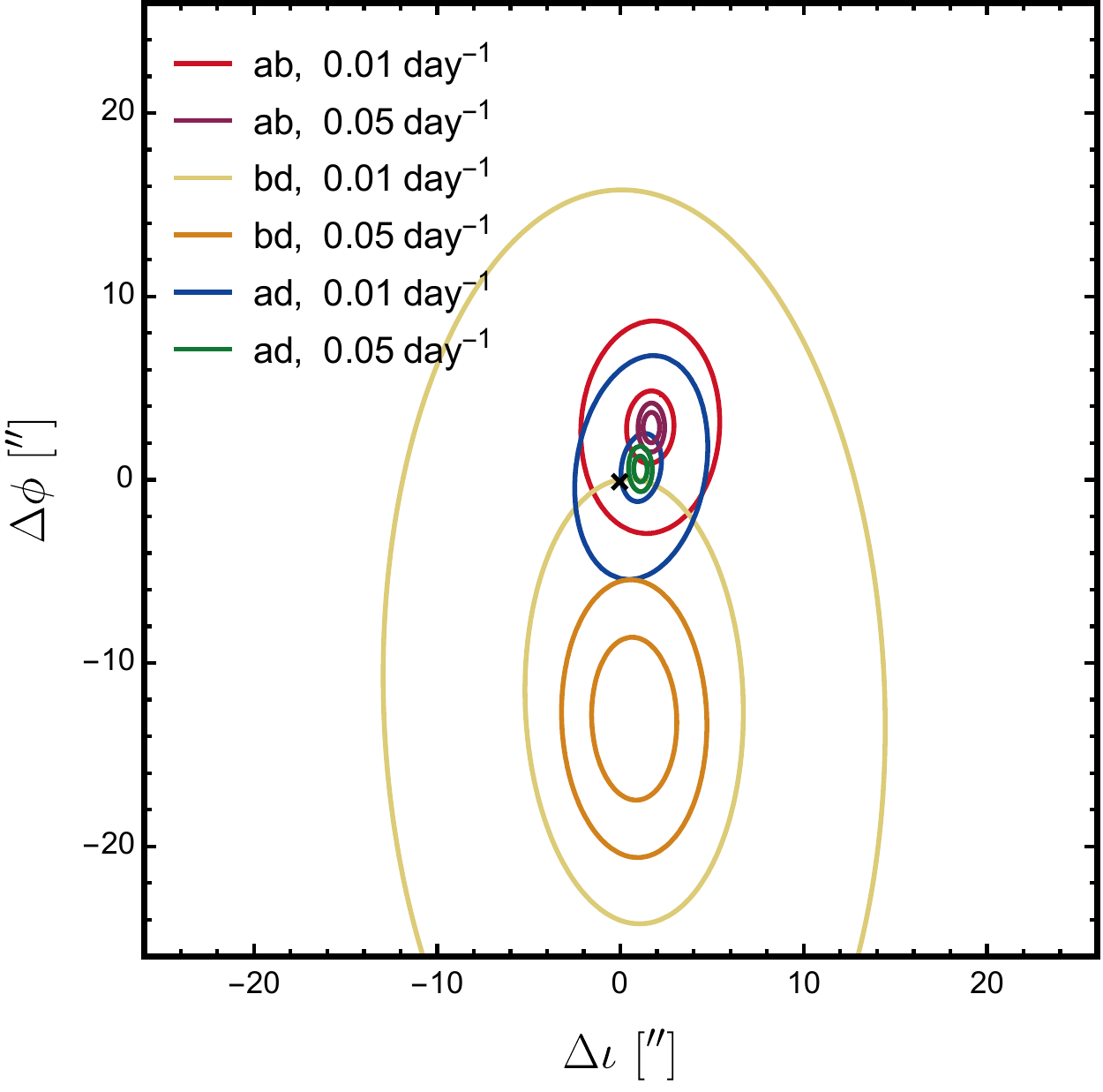}
    \caption{\label{fig:skylocalization} {\it Left}: An example showing simulated time delays between Image $a$ and Image $b$ versus the arrival time of Image $a$. Mock data include a total of 30 repetitions, which are then fit to the five-parameter model of \refeq{5par} neglecting any non-uniform motion transverse to the line of sight. {\it Right}: $1\sigma$ and $2\sigma$ spread of the maximum-likelihood solution for $(\iota,\,\phi)$, from 1000 sets of randomly generated mock data. We consider different choices for the image pair, $(a,b)$, $(b,d)$ and $(a,d)$, and different mean repetition rates. The black cross indicates the true source location.}
  \end{center}
\end{figure}

The left panel of \reffig{skylocalization} gives an example of how the delay between Image $a$ and Image $b$ might exhibit nearly sinusoidal variation among $\sim 30$ repetitions throughout 500 days of observation, which can be well fit by the five-parameter model \refeq{5par}.

To numerically assess the uncertainty in the sky localization, we simulate a large number of mock observations. For each of them, we generate random source repetitions, compute predicted times of arrival, and then measure $(\iota,\phi)$ using \refeq{5par}.  Marginalizing over the nuisance parameters $(\Delta T_{IJ,0},K_I,K_J)$, we infer the precision of localization from the amount of scatter in the best-fit values for $(\iota, \phi)$ around their true values. This is shown in the right panel of \reffig{skylocalization}.

Our results indicate that with the detection of $\sim 5$ ($\sim 25$) repetitions, using time delays for a single pair of images $(a,b)$, the inclination $\iota$ can be localized to $\sim 5''$ ($\sim 0.5''$) at 2$\sigma$. The uncertainty in $\phi$ of similar size. Using an image pair with a shorter time delay (e.g. $(b,d)$ ) leads to worse error-bars. The error-bar can further shrink if delay measurements from several pairs are combined. Since this level of localization is insufficient to resolve images, we have simply assumed the same $(\iota, \phi)$ for all images. 

At a repetition rate of $\sim 0.05\,{\rm day}^{-1}$\footnote{This rate
  is much lower than the intrinsic rate inferred for FRB121102
  \citep{Opperman:2017kql}. However, the observed rate is reduced
  compared to the intrinisic rate, due to limitations from telescope
  sensitivity and observational cadence. The repetition rate we use in
  mock simulations always refers to the observed rate.} with a total
of $\sim 30$ repetitions detected, the statistical uncertainty can be
reduced to $\lesssim 1''$. However, as shown in
\reffig{skylocalization} (right panel) localization will be
systematically biased, in a way that depends on which image pair is being
used. The reason for this bias is that the simple five-parameter model
\refeq{5par} does not capture the (nearly) sinusoidal time-delay
perturbation caused by the transverse projection of the Earth's
orbital motion through \refeq{delTkijperp}, which cannot be predicted
without knowing image separations. As the time delay between the image
pair decreases, the effect of the Earth's orbital motion parallel to
the line of sight also decreases (\refeq{delTkijEarthpara}). By
contrast, the effect from the transverse motion does not vanish, as it
is determined by the image separation. Therefore, using an image pair
of shorter time delay would lead to a larger bias. 

Taking that into account, we may conclude that trustworthy source
localization to about $2-10''$ is achievable, given a pulse
timing precision of a few milliseconds. This agrees with the
anticipation in \refsec{order} that neglecting the Earth's transverse
orbital motion restricts the accuracy of angular localization to about
$10^{-5}$ radian. For many survey telescopes, this level of
localization would still help narrowing down the area on the sky VLBI
follow-ups have to search for \citep{2017arXiv170502998E}.

We add one caveat that millisecond timing precision might be prohibited by severe scattering broadening due to the lens galaxy (see estimates in \refsec{sca-broad-lens}). This is especially problematic at low frequencies $\lesssim 1\,$GHz where many survey instruments will be operating, and for images close to the center of the lens. Scattering broadening is significantly mitigated when observing at higher frequencies $\gtrsim 2\,$GHz (e.g. SKA1-MID \citep{Macquart:2015uea}), but then indirect localization to $10''$ using time delay may not be superior than the instrument's intrinsic resolution. In any case, VLBI follow-ups of a lensed repeater may or may not benefit from this indirect method. Once successfully done, it will help to identify both the host galaxy and the lens.

\section{Probing source motion with VLBI localization}
\label{sec:source}

Localization of a lensed repeater with VLBI, if eventually done, would enable precise angular resolution (e.g. at $5\,{\rm GHz}$ $\sim 5\,{\rm mas}$ for EVN\footnote{\url{http://www.evlbi.org/user_guide/res.html}} and $\sim 1\,{\rm mas}$ for VLBA\footnote{\url{http://www.vlba.nrao.edu/astro/obstatus/2012-01-06/node24.html}}). This should be sufficient to resolve multiple images and measure $\bfx_I$'s (although the true source position $\bfy$ is still unknown without lens modeling). This high level of angular resolution would make it feasible to identify the lens galaxy as well as the lensed source galaxy in the background, following which their redshifts $z_S$ and $z_L$ can be separately determined from optical follow-ups.

Given the high promise of the VLBI technique, we furthermore explore the possibility of probing observable effect of any non-uniform transverse motion of the source, $\delta T^{(k)}_{IJ,s\perp}$ (\refeq{delTkijs}), by accurately subtracting the effect of the Earth's orbital motion $\delta T^{(k)}_{IJ,\oplus\parallel}$ (\refeq{delTkijEarthpara}) and $\delta T^{(k)}_{IJ,\oplus\perp}$ (\refeq{delTkijperp}). Owing to superb VLBI resolution, the latter can be predicted to high precision. The residuals thus carry valuable information about the source.

\subsection{Random emission spots}
\label{sec:rand-spot}

It might be that compact emission spots of radio bursts are hosted by an extended clump of material. Each time one spot within this clump has the right condition for coherent radio emission, a burst is emitted from that spot. As a toy model, let us assume that these emission spots are uniformly distributed and randomly switch on and off within a spherical volume of radius $\mathcal{R}$, then for $\mathcal{R} \gtrsim 1\,{\rm AU}$ detectable time-delay perturbation is induced as the spot of emission switches from one to another. This is only meant to demonstrate what level of displacement in the emission spot can produce detectable signatures. In reality, the geometry of the extended clump and the statistics of emission spots might be completely different.

If we are ignorant of this effect, we may simply use the following model to fit the time-delay data:
\ba
\label{eq:7par}
\Delta T^{(k)}_{IJ} & = & \Delta T_{IJ,0} + \left( K_J\,t^{(k)}_J - K_I\,t^{(k)}_I \right) + \frac{p_\oplus}{c}\,\sin\iota_\star\, \left[ \frac{\cos\left(\phi_\star - \phi_\oplus(t^{(k)}_J)\right)}{1+e_\oplus\,\cos\phi_\oplus(t^{(k)}_J)} -\frac{\cos\left(\phi_\star - \phi_\oplus(t^{(k)}_I)\right)}{1+e_\oplus\,\cos\phi_\oplus(t^{(k)}_I)} \right] \en
&& + \frac{1}{c} \left[ \bfx_I \cdot \bfd_{\oplus}(t^{(k)}_I) - \bfx_J \cdot \bfd_{\oplus}(t^{(k)}_J) \right],
\ea
Note that the location of the optical axis $(\iota_\star,\,\phi_\star)$ is known (it is artificially chosen) while the true source position $\bfy$ is not. We would like to maximize the log likelihood,
\ba
\ln \mathcal{L} & = & - \frac{1}{2\,\sigma^2_\theta}\,\sum^N_{k=1}\,\left[ \left( \bfx^{(k)}_I - \bfx_I \right)^2 + \left( \bfx^{(k)}_J - \bfx_J \right)^2 \right]  - \frac{1}{4\,\sigma^2_w}\,\sum^N_{k=1}\,\Bigg\{ t^{(k)}_J - t^{(k)}_I - \Delta T_{IJ,0} - \,\left( K_J\,t^{(k)}_J - K_I\,t^{(k)}_I \right)  \en
&&  - \frac{p_\oplus}{c}\,\sin\iota_\star\, \left[ \frac{\cos\left(\phi_\star - \phi_\oplus(t^{(k)}_J)\right)}{1+e_\oplus\,\cos\phi_\oplus(t^{(k)}_J)} -\frac{\cos\left(\phi_\star - \phi_\oplus(t^{(k)}_I)\right)}{1+e_\oplus\,\cos\phi_\oplus(t^{(k)}_I)} \right] - \frac{1}{c} \left[ \bfx_I \cdot \bfd_{\oplus}(t^{(k)}_I) - \bfx_J \cdot \bfd_{\oplus}(t^{(k)}_J) \right]  \Bigg\}^2,
\ea
with respect to seven free parameters $(\bfx_I, \bfx_J, \Delta T_{IJ,0}, K_I, K_J)$. Here $\bfx^{(k)}_I$ is the measured angular position of the $I^{\rm th}$ image in the $k^{\rm th}$ repetition. We assume that with VLBI $\bfx^{(k)}_I$'s are measured with a standard deviation $\sigma_\theta$. The nuisance parameters $(\Delta T_{IJ,0}, K_I, K_J)$ are introduced to account for the constant and the linearly-drifting part of the time delay.

\begin{figure}[t]
  \begin{center}
    \includegraphics[scale=0.6]{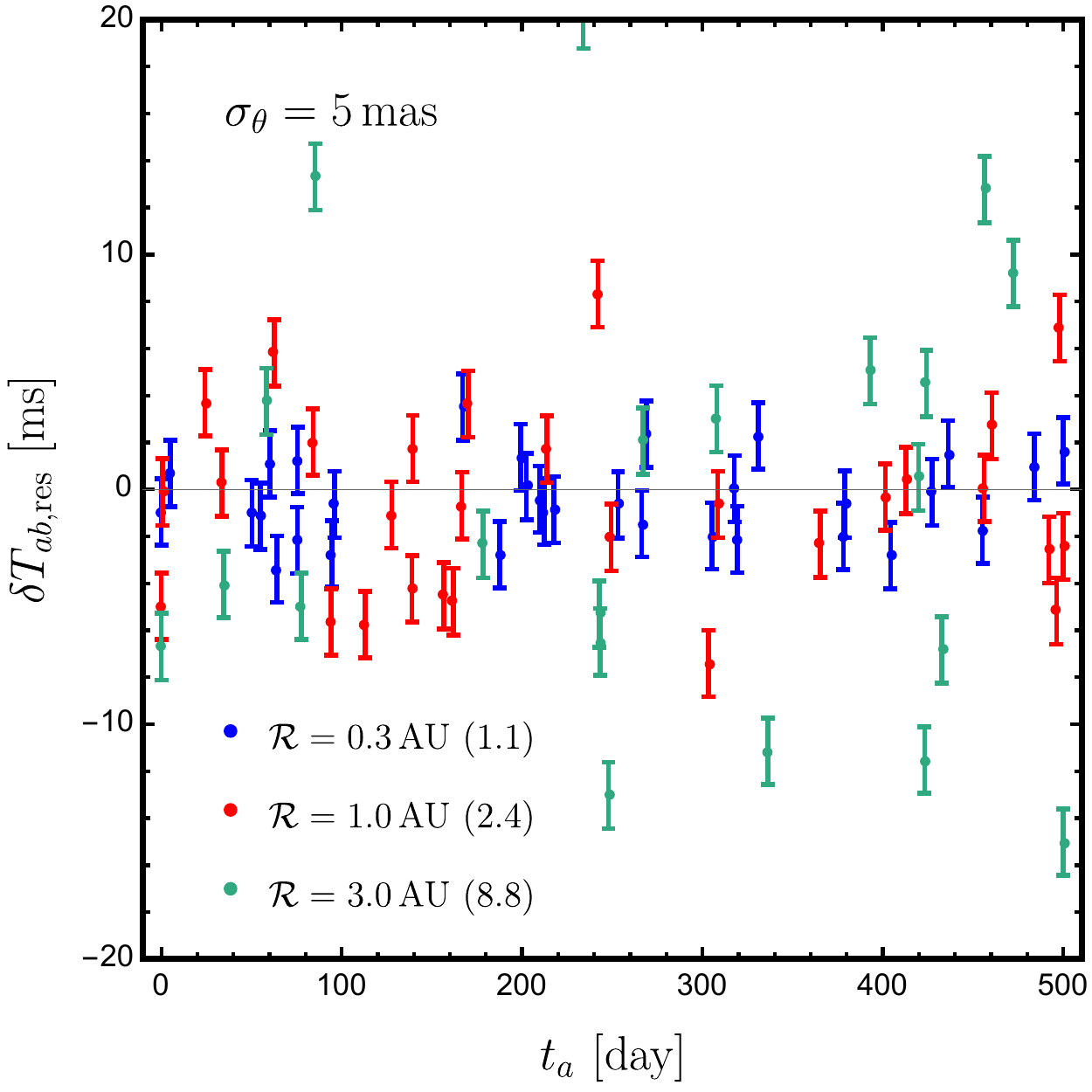}\qquad
    \includegraphics[scale=0.58]{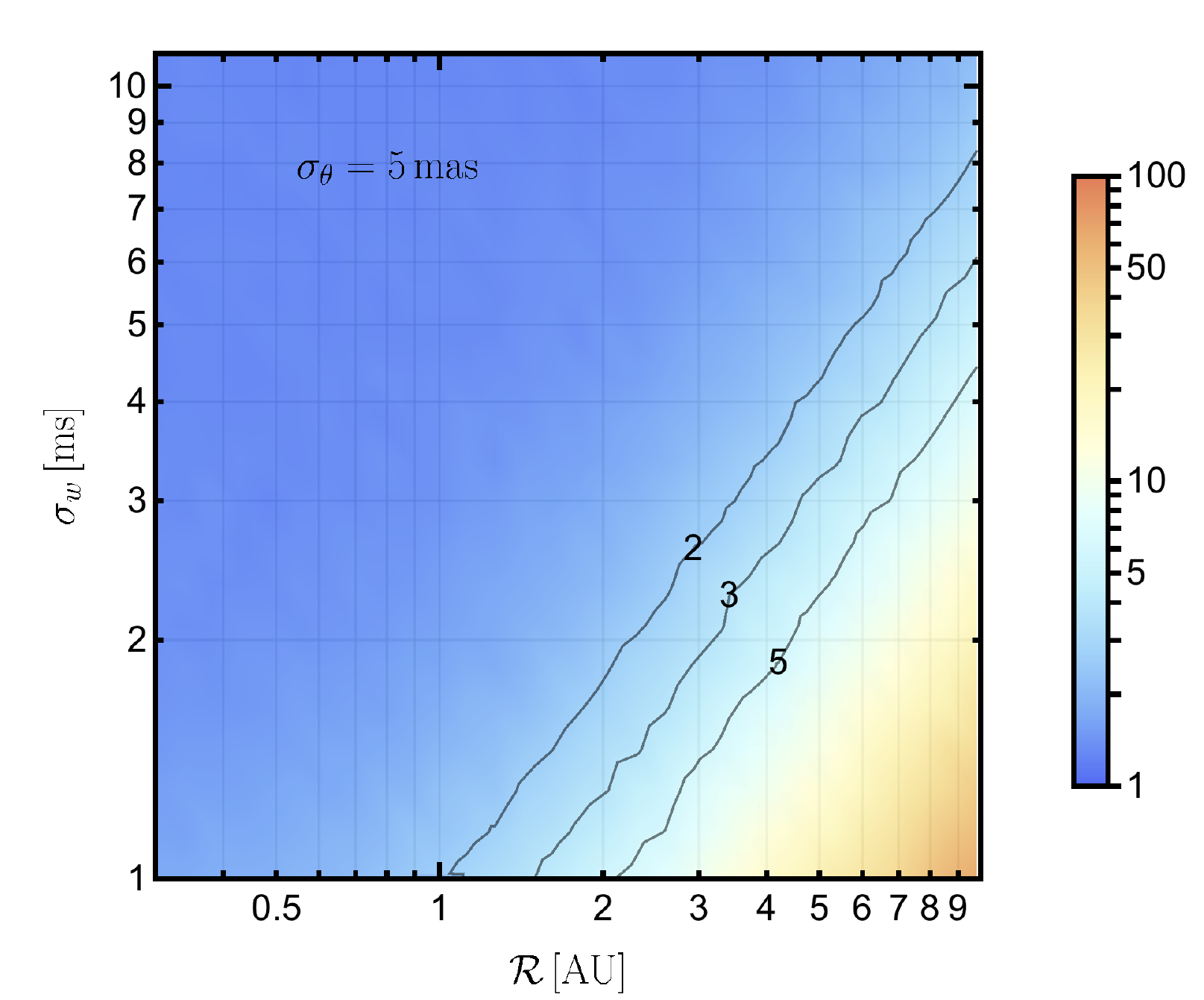}
    \caption{\label{fig:residue} {\it Left:} Simulated time-delay residuals after fitting to \refeq{7par}. Error-bars are derived from a constant timing error $\sigma_w = 1\,{\rm ms}$. We consider different radii of scatter $\mathcal{R}$. The number in the parenthesis indicates the $\chi^2/{\rm d.o.f.}$ for the model \refeq{7par}. {\it Right:} $\chi^2/{\rm d.o.f.}$ as a function of $\mathcal{R}$ and $\sigma_w$, with each pixel computed by averaging over 50 independent mock simulations. Contours of $\chi^2/{\rm d.o.f.} = 2,\,3,\,5$ are drawn. In both panels, we assume a repetition rate $0.05\,{\rm day}^{-1}$ and $\sigma_\theta = 5\,{\rm mas}$.}
  \end{center}
\end{figure}

In the left panel of \reffig{residue}, we show examples of residuals after fitting the mock time-delay measurement for the image pair $(a,b)$. We assume a repetition rate $0.05\,{\rm day}^{-1}$ and an angular resolution $\sigma_\theta = 5\,{\rm mas}$. We consider different radii for the extended clump $\mathcal{R} = 0.1\,{\rm AU}, \,1\,{\rm AU}, \,3\,{\rm AU}$. For small radii $\mathcal{R} \ll 1\,{\rm AU}$, \refeq{7par} provides a good fit to the noisy data, giving a $\chi^2$ per degree of freedom close to unity. For larger radii $\mathcal{R} \sim 3\,{\rm AU}$, \refeq{7par} gives a $\chi^2$ per degree of freedom that is way too high, suggesting that time-delay perturbations caused by randomized emission spots are resolved. To see which regions in the $\mathcal{R}-\sigma_w$ parameter space can be probed by timing measurements, we show in the right panel of \reffig{residue} the typical $\chi^2$ per degree of freedom as a function of both $\mathcal{R}$ and $\sigma_w$. Since the smooth model \refeq{7par} does not produce stochastic fluctuations in the time delay, the conclusion will be robust even for poor resolution $\sigma_\theta = 100\,{\rm mas}$. Again, the significance of the measurement can be further improved by jointly fitting the time delays for several image pairs.

\subsection{Orbital motion}

Another possibility is that the compact source orbits around another mass. Orbital motion projected onto the plane of the sky perturbs the lensing time delay via \refeq{delTkijs_mutual}. For a simple but concrete example, consider a circular orbit for the source\footnote{The method can be easily generalized to the case of an eccentric orbit, which might be relevant for a young neutron star in a binary system.}, whose normal is at an angle $\iota_s$ ($ 0 \leq \iota_s <\pi$) to the optical axis and its projection onto the plane of the sky has a major axis at an angle $\phi_s$ ($ 0 \leq \phi_s <\pi$) to the first coordinate axis.

The oscillatory part of the source's transverse displacement vector (barring constant motion), in its components, is given by
\ba
\hspace{-1cm}
\bfd_s(t_s) = a_s\,&& \Big( \cos\iota_s\,\cos\phi_s\,\cos\left( \Omega_s\,t_s - \varphi_s \right) + \sin\phi_s\,\sin\left( \Omega_s\,t_s - \varphi_s \right), \en
&& \cos\iota_s\,\sin\phi_s\,\cos\left( \Omega_s\,t_s - \varphi_s \right) - \cos\phi_s\,\sin\left( \Omega_s\,t_s - \varphi_s \right) \Big),
\ea
where $a_s > 0$ is the orbital radius, $\Omega_s > 0$ is the source-frame orbital (angular) frequency, and $\varphi_s$ ($ 0 \leq \varphi_s < 2\pi$) is the orbital phase. For a single pair of images, we propose the following timing model
\ba
\label{eq:9par}
\Delta T^{(k)}_{IJ} & = & \Delta T_{IJ,0} + \left( K_J\,t^{(k)}_J - K_I\,t^{(k)}_I \right) + \frac{p_\oplus}{c}\,\sin\iota_\star\, \left[ \frac{\cos\left(\phi_\star - \phi_\oplus(t^{(k)}_J)\right)}{1+e_\oplus\,\cos\phi_\oplus(t^{(k)}_J)} -\frac{\cos\left(\phi_\star - \phi_\oplus(t^{(k)}_I)\right)}{1+e_\oplus\,\cos\phi_\oplus(t^{(k)}_I)} \right] \en
&& + \frac{1}{c} \left[ \bfx_I \cdot \bfd_{\oplus}(t^{(k)}_I) - \bfx_J \cdot \bfd_{\oplus}(t^{(k)}_J) \right] + (1+z_L)\,\frac{D_L}{c\,D_{LS}}\,\left[ A_{IJ}\,\cos\left( \Omega_s\,t^{(k)}_s \right) + B_{IJ}\,\sin\left( \Omega_s\,t^{(k)}_s \right) \right],
\ea
where we define the source-frame time $t_s$ using $t^{(k)}_s = t^{(k)}_I/(1+z_S)$. The last two terms describe a sinusoidal perturbation to the time delay due to the source's orbital motion, parametrized by a source-frame frequency $\Omega_s$. If $z_S$ is not known, then only the redshifted orbital frequency $\Omega_s/(1+z_S)$ is measurable. The two coefficients $A_{IJ}$ and $B_{IJ}$ are dependent on the image separation as well as the orbital parameters,
\ba
\label{eq:solvesourcepar}
\hspace{-0.5cm}
\begin{cases}
& A_{IJ} = a_s\,\left[\left( x_{I1} - x_{J1} \right) \,\left( \cos\iota_s\,\cos\phi_s\,\cos\varphi_s - \sin\phi_s\,\sin\varphi_s \right) + \left( x_{I2} - x_{J2} \right) \,\left( \cos\iota_s\,\sin\phi_s\,\cos\varphi_s + \cos\phi_s\,\sin\varphi_s \right) \right], \\
& B_{IJ} = a_s\,\left[\left( x_{I1} - x_{J1} \right) \,\left( \cos\iota_s\,\cos\phi_s\,\sin\varphi_s + \sin\phi_s\,\cos\varphi_s \right) + \left( x_{I2} - x_{J2} \right) \,\left( \cos\iota_s\,\sin\phi_s\,\sin\varphi_s - \cos\phi_s\,\cos\varphi_s \right) \right].
\end{cases}
\ea
A single pair of images is sufficient to infer $\Omega_s$ (assume redshifts are known). However, at least two pairs with linearly independent image separation vectors are required to separately determine the other orbital parameters $a_s$, $\iota_s$, $\phi_s$ and $\varphi_s$. For three images forming two pairs $(I,J)$ and $(I,K)$, whose image separation vectors are in general not collinear, we can use a timing model containing 11 nuisance parameters plus 5 parameters that are related to source motion,
\ba
\left( \bfx_I,\,\bfx_J,\,\bfx_K,\,\Delta T_{IJ,0},\,\Delta T_{IK,0},\,K_I,\,K_J,\,K_K;\,\, \Omega_s,\,A_{IJ},\,B_{IJ},\,A_{IK},\,B_{IK} \right).
\ea
From the last 4 parameters $\left( A_{IJ},\,B_{IJ},\,A_{IK},\,B_{IK} \right)$ we can solve for $\left( a_s,\,\iota_s,\,\phi_s,\,\varphi_s \right)$. The corresponding likelihood reads
\ba
\label{eq:10par-lnL}
\ln\mathcal{L} & = & - \frac{1}{2\,\sigma^2_\theta}\,\sum^N_{k=1}\,\left[  \left( \bfx^{(k)}_I - \bfx_I \right)^2 + \left( \bfx^{(k)}_J - \bfx_J \right)^2 +  \left( \bfx^{(k)}_K - \bfx_K \right)^2\right] \en
&& - \frac{1}{6\,\sigma^2_w}\,\sum^N_{k=1}\,\left[\begin{array}{cc}
\delta T_{IJ}^{(k)}, & \delta T_{IK}^{(k)}\end{array}\right]\, \left[\begin{array}{cc}
2 & -1\\
-1 & 2
\end{array}\right]\, \left[\begin{array}{c}
\delta T_{IJ}^{(k)}\\
\delta T_{IK}^{(k)}
\end{array}\right],
\ea
where the residual for a given image pair $(I,J)$ is given by
\ba
\label{eq:10par-delT}
\delta T^{(k)}_{IJ}  & = & t^{(k)}_J - t^{(k)}_I - \Delta T_{IJ,0} - \,\left( K_J\,t^{(k)}_J - K_I\,t^{(k)}_I \right)  - \frac{p_\oplus\,\sin\iota_\star}{c}\, \left[ \frac{\cos\left(\phi_\star - \phi_\oplus(t^{(k)}_J)\right)}{1+e_\oplus\,\cos\phi_\oplus(t^{(k)}_J)} -\frac{\cos\left(\phi_\star - \phi_\oplus(t^{(k)}_I)\right)}{1+e_\oplus\,\cos\phi_\oplus(t^{(k)}_I)} \right] \en
&& - \frac{1}{c} \left[ \bfx_I \cdot \bfd_{\oplus}(t^{(k)}_I) - \bfx_J \cdot \bfd_{\oplus}(t^{(k)}_J) \right] - (1+z_L)\,\frac{D_L}{c\,D_{LS}}\,\left[ A_{IJ}\,\cos\left( \Omega_s\,t^{(k)}_s \right) + B_{IJ}\,\sin\left( \Omega_s\,t^{(k)}_s \right) \right].
\ea
Since the fitted value for $\varphi_s$ artificially depends on the choice of zero-point for $t_s$, it is of limited physical interest and is essentially another nuisance parameter.

\begin{figure}[t]
  \begin{center}
    \includegraphics[scale=0.63]{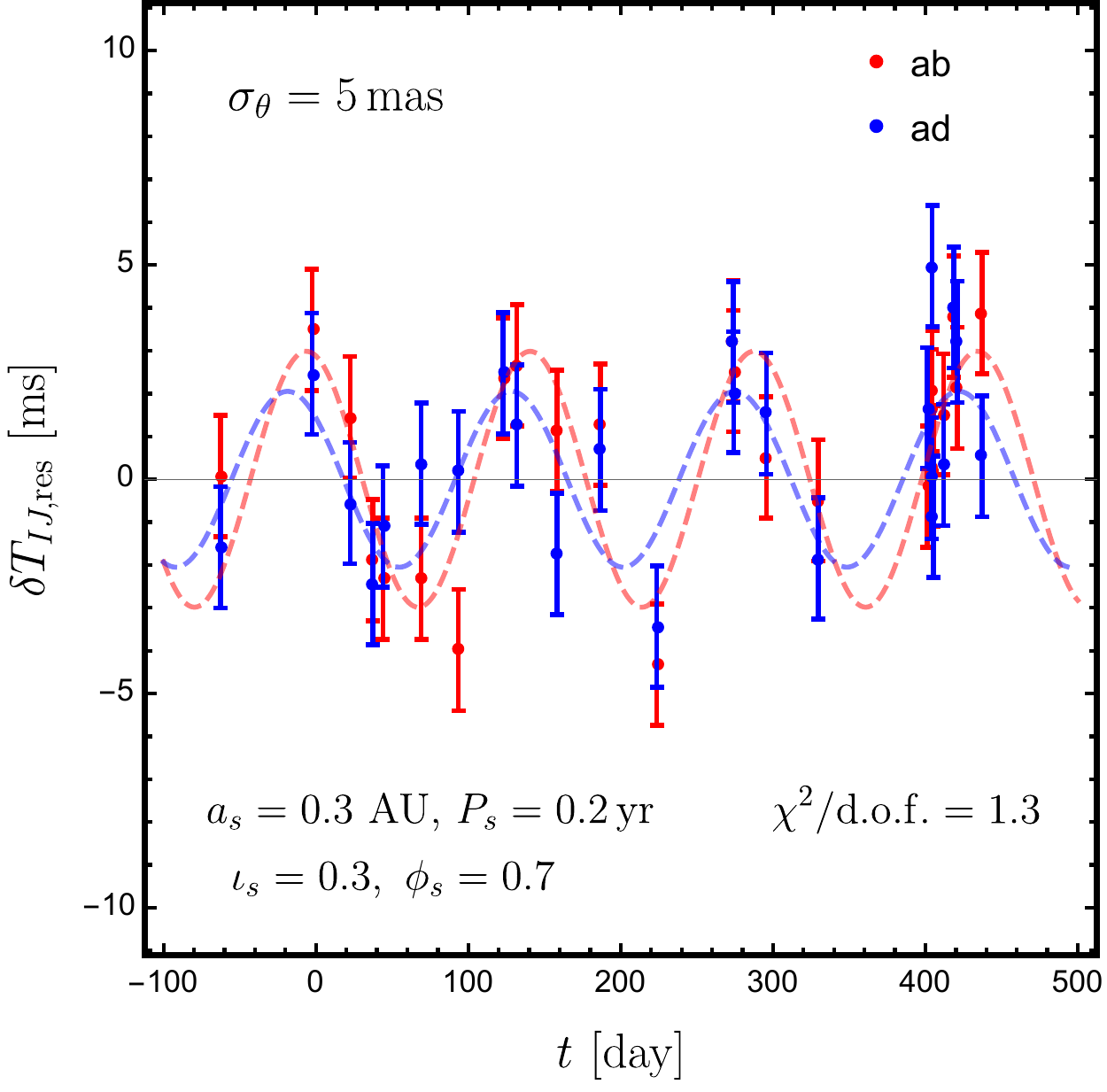}\qquad
    \includegraphics[scale=0.63]{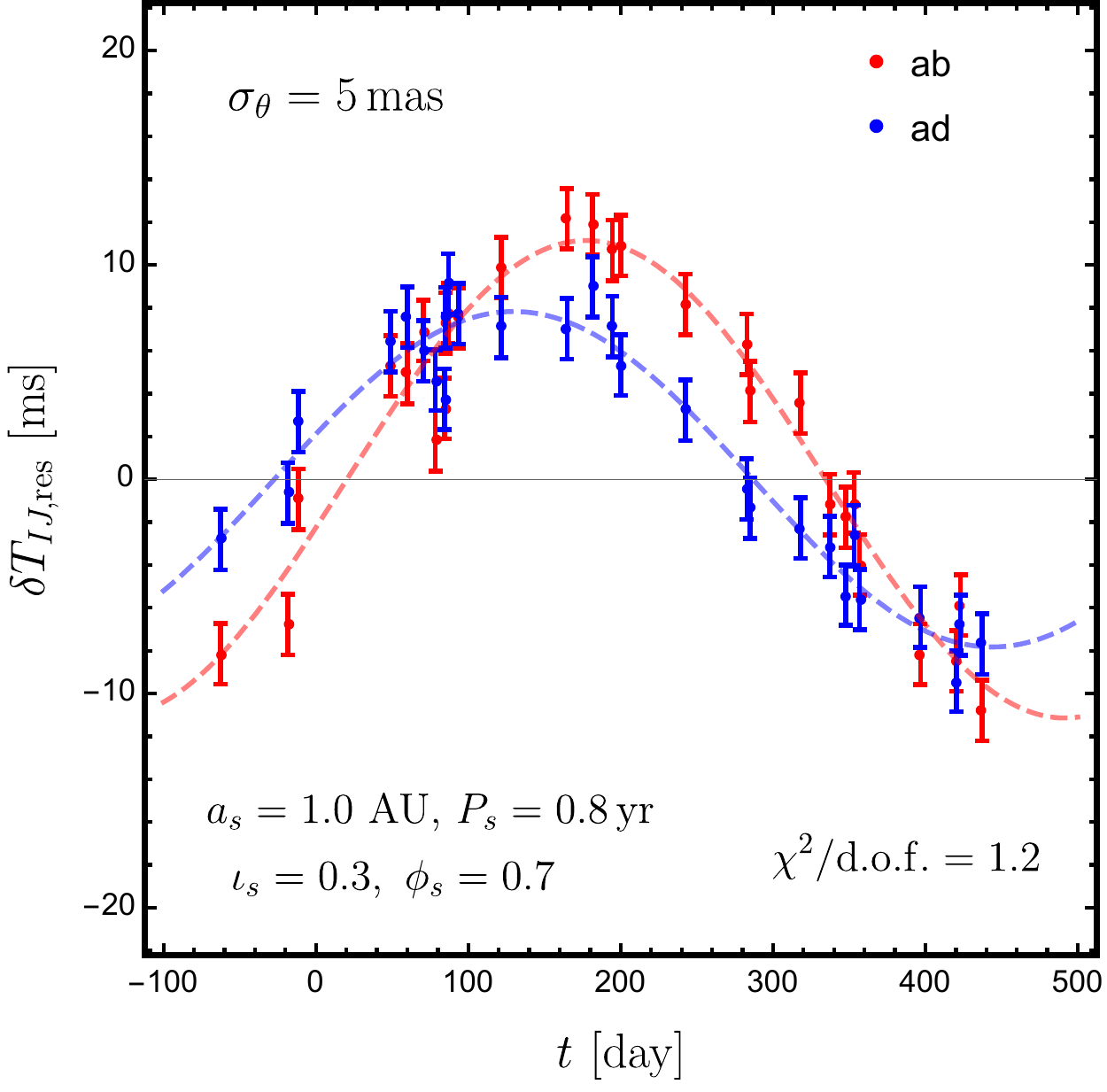}
    \caption{\label{fig:source-orbital-residual}. Two examples of mock time-delay residuals due to the source's orbital motion and the corresponding best fits using the two-image-pair model \refeqs{10par-lnL}{10par-delT}. Two image pairs $(a,\,b)$ and $(a,\,d)$ are used. Error-bars are derived from a constant timing error $\sigma_w = 1\,{\rm ms}$ and a constant localization error $\sigma_\theta = 5\,{\rm mas}$. A repetition rate $0.05\,{\rm day}^{-1}$ is assumed for an observation spanning 500 days. For the panel on the left, maximum-likelihood estimation gives $a_s = 0.31\,{\rm AU}$, $P_s = (2\,\pi)/\Omega_s = 0.20\,{\rm yr}$, $\iota_s = 0.78$ and $\phi_s = 0.15$; for the panel on the right, maximum-likelihood estimation gives $a_s = 1.1\,{\rm AU}$, $P_s = (2\,\pi)/\Omega_s = 0.86\,{\rm yr}$, $\iota_s = 0.76$ and $\phi_s = 0.32$.}
  \end{center}
\end{figure}

In \reffig{source-orbital-residual}, we give examples of how the time-delay residuals between two pairs of images would look if the source has an orbital motion on the timescale of $\lesssim 1\,{\rm yr}$ with a radius $\sim 0.3 - 1\,{\rm AU}$, which is typical for a stellar companion. A repetition rate $0.05\,{\rm day}^{-1}$ is assumed throughout a $500$-day observation. It can be seen that \refeq{9par} provides a good fit to the mock data, with source parameters solved using \refeq{solvesourcepar}. We found that for timing accuracy $\sigma_w = 1\,{\rm ms}$ the orbital radius $a_s$ and the orbital period $P_s = (2\,\pi)/\Omega_s$ can be recovered with good accuracy, while the orientation angles $\iota_s$ and $\phi_s$ are subject to large uncertainty. However, $a_s$ and $P_s$ are of major astrophysical interest here as they imply the mass of the companion.

\begin{figure}[t]
  \begin{center}
    \includegraphics[scale=0.63]{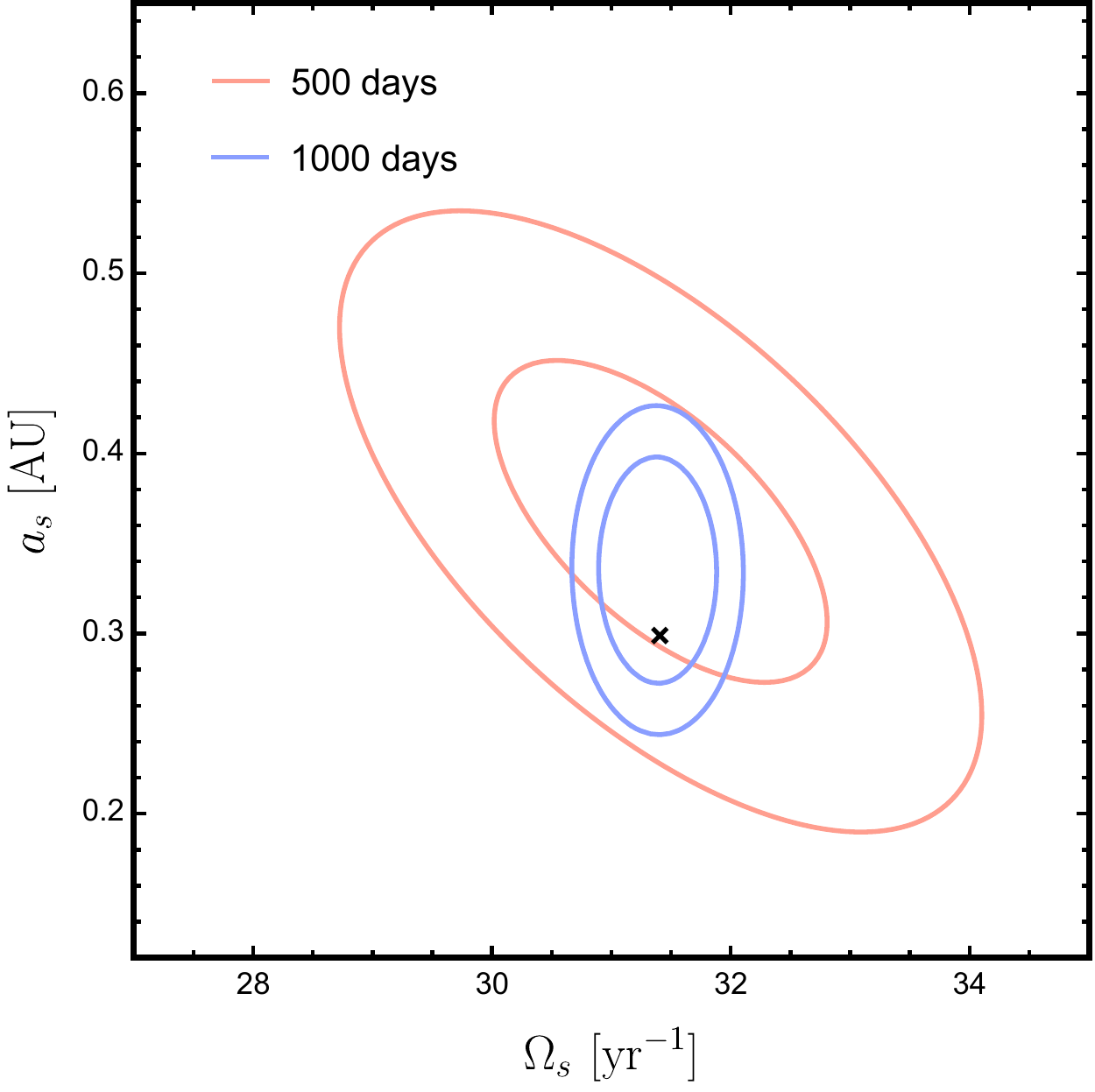}\qquad
    \includegraphics[scale=0.63]{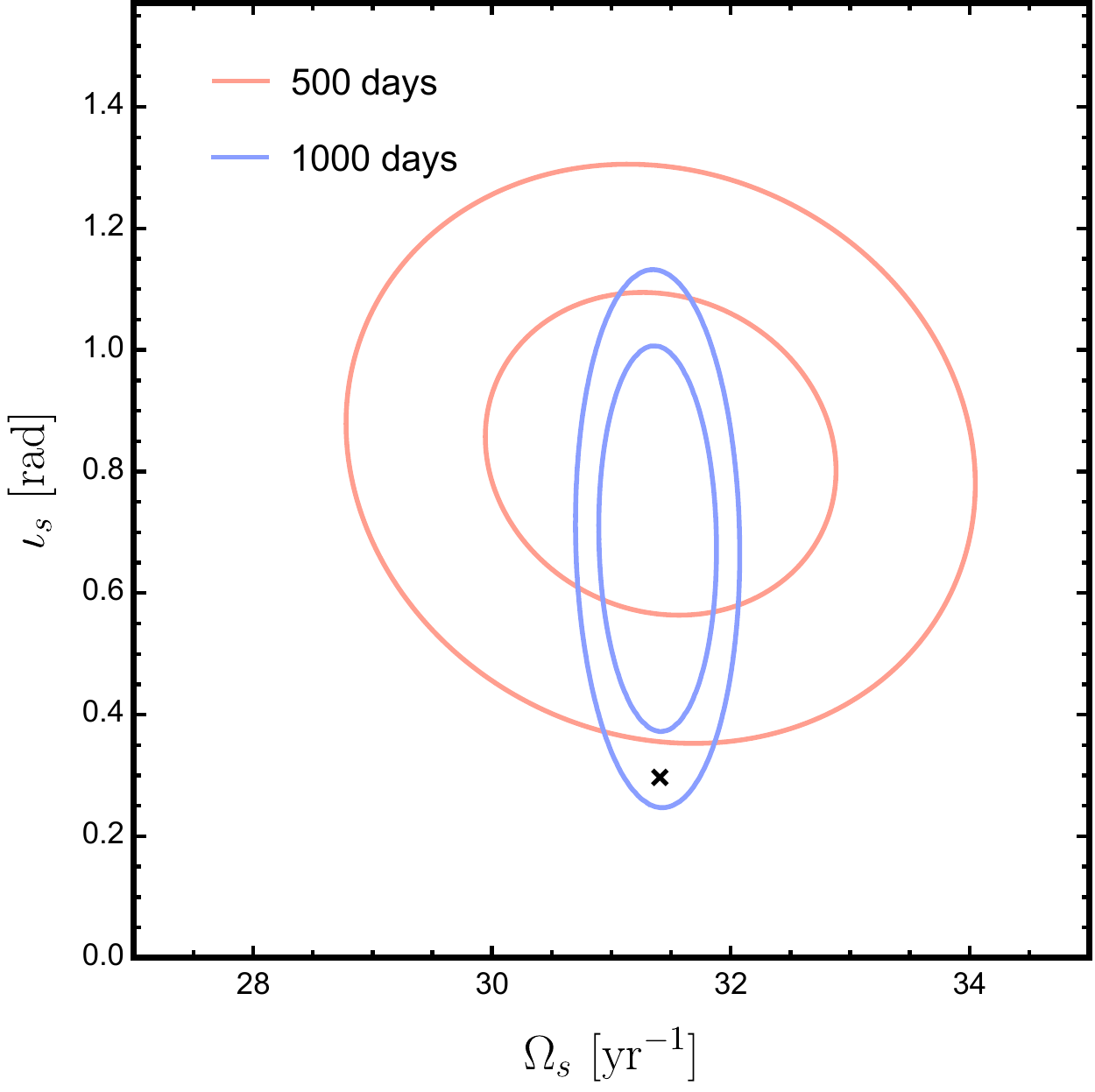} \\ 
    \vspace{0.5cm} \hspace{0.05cm}
        \includegraphics[scale=0.64]{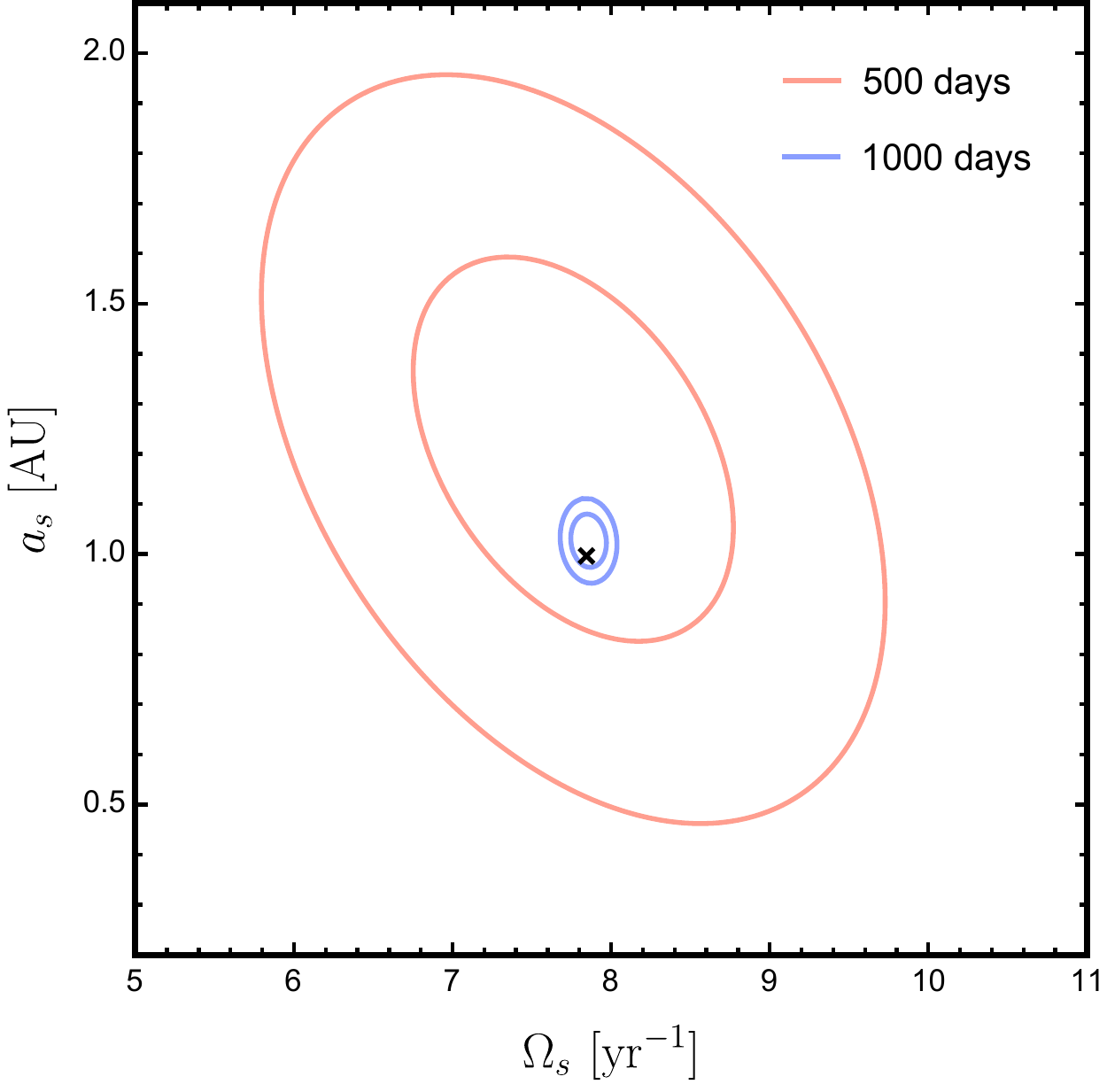}\quad\,\,
    \includegraphics[scale=0.64]{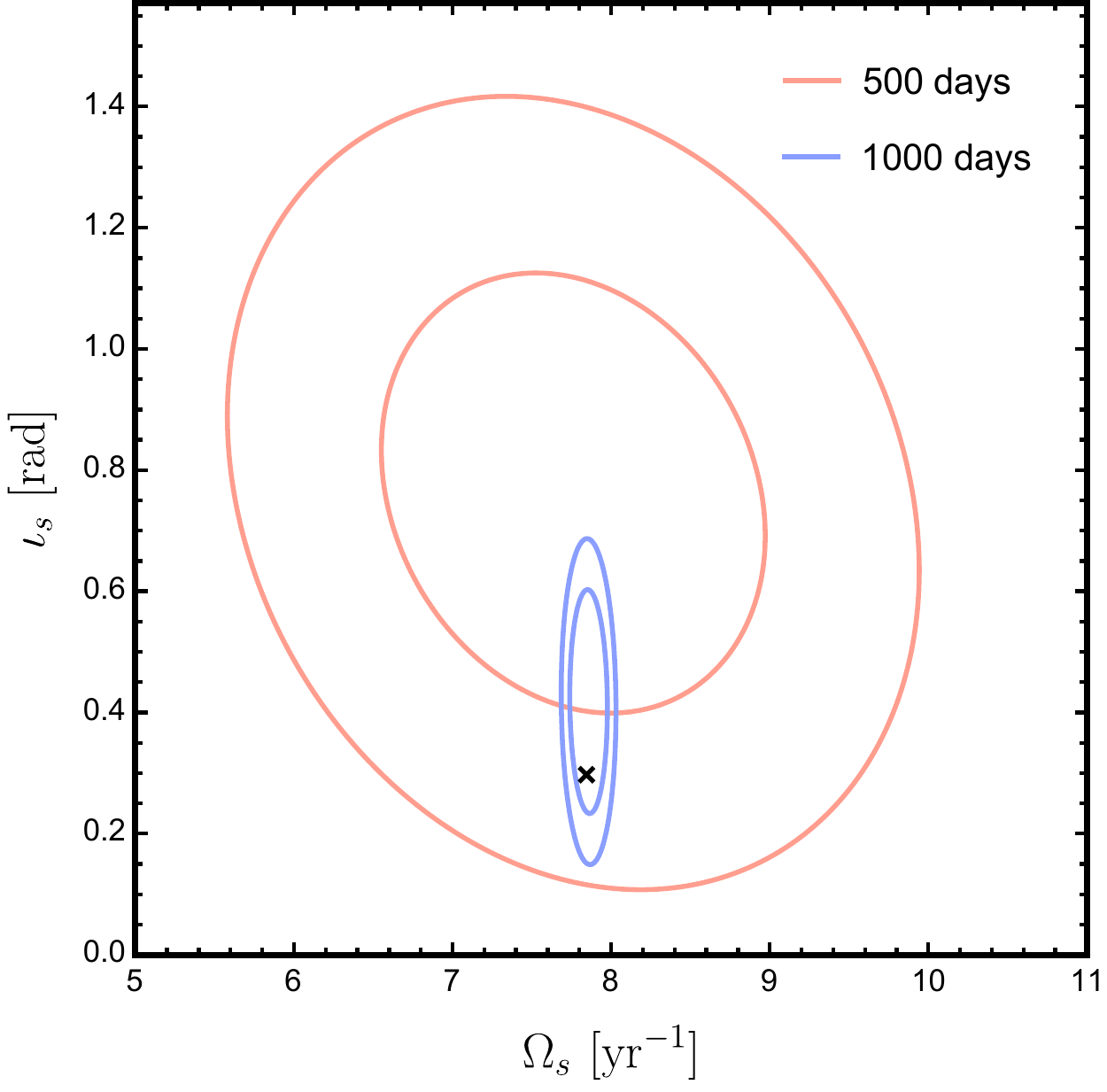} \\
    \caption{\label{fig:source-orbital-par} 1$\sigma$ and 2$\sigma$ joint spread in the maximum-likelihood estimation for the source's orbital frequency $\Omega_s$, semi-major axis $a_s$ and inclination $\iota_s$, derived by simulating 500 independent observations. Taking the strong lensing configuration of \reffig{lens}, we use time delays between two image pairs $(a,\,b)$ and $(a,\,d)$ simultaneously. We assume a repetition rate $0.05\,{\rm day}^{-1}$, timing accuracy $\sigma_w = 1\,{\rm ms}$, and image localization accuracy $\sigma_\theta = 5\,{\rm mas}$. Two different lengths of observation are in comparison: $500\,{\rm days}$ (orange) and $1000\,{\rm days}$ (blue). The upper panels are for the case $P_s = 0.2\,{\rm yr}$ and $a_s = 0.3\,{\rm AU}$, and the lower panels are for the case $P_s = 0.8\,{\rm yr}$ and $a_s = 1.0\,{\rm AU}$. In all cases we fix $\iota_s = 0.3$, $\phi_s = 0.7$ and $\varphi_s = 0.9$. In each panel, a black cross marks the correct values for the source orbital parameters.}
  \end{center}
\end{figure}

In \reffig{source-orbital-par}, we demonstrate how well source parameters can be typically measured from time-delay residuals, by simulating a large number of mock observations. In particular, we focus on three parameters of foremost astrophysical interest: the orbital frequency $\Omega_s$, the radius $a_s$, and the inclination $\iota_s$. The orbital frequency $\Omega_s$ in general can be recovered with fairly good accuracy. An observational span of 500 days is sensitive to probe orbital period on the order of $0.1 - 1\,{\rm yr}$. An even longer observational span would significantly improve the measurement for the case of longer orbital period $P_s = (2\,\pi)/\Omega_s \gtrsim 1\,{\rm yr}$. Typically, with a timing accuracy $\sigma_w \sim 1\,{\rm ms}$, the orbital radius is measurable if $a_s \gtrsim 0.1\,{\rm AU}$. Smaller orbital amplitudes produce time-delay perturbations that are too small to be recognizable. However, as shown in \reffig{source-orbital-par}, it is more difficult to determine the orbital inclination $\iota_s$ accurately. 

In summary, source binary motion with separation on the order of
$\gtrsim 0.1 - 1\,{\rm AU}$ is detectable under reasonable assumptions
about the quality of VLBI observation of a lensed repeater. Generally
speaking, three factors can improve the measurement of the orbital
motion: (1) better timing accuracy for individual bursts; (2) longer
observational span; (3) detection of more repetitions.

For another plausible situation, if the source closely orbits a massive black hole, acceleration transverse to the line of sight will induce a quadratic deviation from simple linear drift in the lensing time delay, which should be detectable if acceleration generates additional transverse displacement that accumulates to $\gtrsim 0.1\,{\rm AU}$ over the observational time span. For an order-of-magnitude estimate, the additional displacement is given by
\ba
\frac{G\,M_{\rm BH}}{R^2}\,\left(\frac{T_{\rm obs}}{1+z_S} \right)^2 \approx 0.6\,{\rm AU}\,\left( \frac{10^6\,M_\odot}{M_{\rm BH}} \right)\,\left( \frac{0.1\,{\rm pc}}{R} \right)^2\,\left( \frac{T_{\rm obs}}{5\,{\rm yr}} \right)^2\,\left( \frac{2}{1+z_S} \right)^2,
\ea 
where $M_{\rm BH}$ is the black hole mass and $R$ is the typical distance to the black hole. In this case, the orbital period is roughly
\ba
2\,\pi\,\left( \frac{R^3}{G\,M_{\rm BH}} \right)^{1/2} \approx 3000\,{\rm yr}\,\left( \frac{R}{0.1\,{\rm pc}} \right)^{3/2}\,\left( \frac{10^6\,M_\odot}{M_{\rm BH}} \right)^{1/2},
\ea
much longer than the observational time span.

\section{Discussion}
\label{sec:discussion}

Our method is based on the assumption that time of arrival can be
measured to an accuracy $\sim1\,{\rm ms}$. However, a received
(de-dispersed) pulse could still be broadened, due to either
gravitational microlensing by stars in the lens galaxy, or scattering
by the inhomogeneous ISM in the lens galaxy. 

To address these issues, in \refsec{micro-broad}, we first discuss the
monochromatic effect of microlensing on pulse broadening and show that it is most
likely unimportant. Then in \refsec{sca-broad-lens}, we estimate
scattering broadening by the ISM of the lens galaxy, which may adversely affect detection and timing at low frequencies. Then in \refsec{host-scatter}, we discuss angular broadening of the source due to scattering in the host galaxy.

\subsection{Broadening due to microlensing}
\label{sec:micro-broad}

Taking the same example ($z_L=0.5$ and
$z_S=1$) as in \refsec{simulation}, the critical surface mass density is given by 
$\Sigma_{\rm crit} = 
c^2\,D_{\rm S}/(4\,\pi\, G\, D_L\,D_{LS})\simeq 3000\, M_{\odot} \rm\
pc^{-2}$. We define the optical depth for microlensing as $\sigma_* =
\Sigma_*/\Sigma_{\rm crit}$, where $\Sigma_*$ is the surface mass
density of stars (including stars of all evolution stages and compact
objects). If the image is behind the outskirts of the lens galaxy, we
have $\sigma_*\ll 1$ and at most one star may cause microlensing. Then the
delayed time between the two micro-images is given by
\begin{equation}
  \label{eq:4}
  \Delta t_{\rm ML, thin}\simeq {2\,r_E^2\over D_{\rm
      L}c}\simeq 7\times 10^{-3}\rm\ ms
\end{equation}
where
\begin{equation}
  \label{eq:2}
  r_E = \left(\frac{4\,G\,M_*}{c^2}\,\frac{D_{L}\,D_{LS}}{D_{S}}\right)^{1/2}\simeq
  2\times10^{16}\rm\ cm
\end{equation}
is the Einstein radius and we have taken an average stellar mass $M_* = 0.4\,M_\odot$. On the other hand, if radio waves pass
within the central few kpc of the lens galaxy, microlensing optical depth may reach order unity. Then 
multiple microlenses may be strongly coupled and $90\%$ of the flux from
numerous micro-images spread out to a typical
angular scale of \citep{1986ApJ...306....2K}
\begin{equation}
  \label{eq:1}
\Delta \theta_{\rm ML, \pm} = 3\,{r_E \over D_L} {\sigma_*^{1/2}
    \over |1 - \sigma_* \pm \gamma|},
\end{equation}
where $\gamma$ is the local macrolensing shear and ``$\pm$''
correspond to the two principal directions that diagonalize the
macrolensing distortion matrix $\partial \b{y}/\partial\b{x}$. The surface
brightness beyond this angular scale decreases rapidly as distance to
the fourth power. The temporal broadening of the FRB is given by the
maximum time delay between the micro-images
\begin{equation}
  \label{eq:3}
  \Delta t_{\rm ML,thick}\simeq 3\,[\mathrm{max}(\Delta \theta_{\rm ML, +}, \Delta
  \theta_{\rm ML, -})]^2\,D_L / c\simeq 10\,\sigma_*
  \mathrm{max} \left[{1\over (1 - \sigma_* + \gamma)^2}, {1\over (1 -
    \sigma_* - \gamma)^2}\right] \times \Delta t_{\rm ML,thin}.
\end{equation}
According to \refeq{3}, only near caustics $\sigma_*+ \gamma$ or $\sigma_* - \gamma$ may become close to 1 and
then the macro-image can be broadened by more than $\sim\,$ms. Therefore, significant microlensing broadening is expected to occur only rarely.

\subsection{Scattering broadening by the lens galaxy}
\label{sec:sca-broad-lens}

Next, we consider scattering broadening due to
the ISM of the lens galaxy. The observed wavefront of a point source
at cosmological distances is subject to phase fluctuations on the lens
plane due to turbulent electron density fluctuations, the power
spectrum of which follows a power-law between some inner length-scale
$l_0\lesssim 100$~km \citep{1990ApJ...353L..29S, 1995ApJ...443..209A} and some outer length-scale $L_0\gtrsim 100$~pc
\citep{1995ApJ...443..209A}. If we assume a Kolmogorov spectrum 
(power-law index $\beta =  11/3$) and a spiral galaxy like the Milky
Way, the amplitude of the 
turbulence per unit length is given by \citep{1995ApJ...443..209A}
\begin{equation}
  \label{eq:11}
  C_{N}^2 ={(\beta-3) n_{\rm rms}^2  L_0^{3-\beta}
    \over 2(2\pi)^{4-\beta}} \simeq (8.5\times10^{-5}\ 
  \mathrm{m}^{-20/3})\, n_{\rm rms,-1.5}^2
L_{\rm 0,2}^{-2/3},
\end{equation}
where
$n_{\rm rms} = \,n_{\rm rms,-1.5}\,10^{-1.5} \rm cm^{-3}\equiv \langle \delta
n_{\rm e}^2\rangle^{1/2}$ is the root-mean-square (rms)
electron density and $L_0= L_{\rm 0,2}\,100\rm\ pc$ is the outer scale. The 
scattering measure (SM; strength of scattering) is given by
the turbulence amplitude $C_{N}^2$ multiplied by the path length
through the lens galaxy $L_{\rm gal} = L_{\rm gal, kpc}\rm\ kpc$,
i.e. $\mathrm{SM} \simeq 
(8.5\times10^{-5} \ \mathrm{kpc\ m}^{-20/3})\, n_{\rm rms,-1.5}^2\,
L_{\rm 0,2}^{-2/3} L_{\rm gal,kpc}$. Note that typical lines of sight
perpendicular to the Milky Way disk in the solar neighborhood have
$\mathrm{SM}\sim 10^{-4}\,\mathrm{kpc\ 
  m}^{-20/3}$ and the mean number density (pulsars' DMs divided by their
distances) $\bar{n}_{\rm e}\sim 10^{-1.5}\rm\,cm^{-3}$
\citep{2003astro.ph..1598C}. In fact, the lens galaxy is more likely a giant elliptical
with little star formation but significant gas content dominated by hot 
ionized medium ($T\sim1\rm\,keV$). Compared to the Milky Way,
the gas density of a giant elliptical is typically lower $n_{\rm e}\sim
10^{-2}\rm\,cm^{-3}$ but the path length is longer $L_{\rm 
gal}\gtrsim10\rm\,kpc$ \citep{2003ARA&A..41..191M}. There have been
observational evidences for angular broadening of strongly lensed
extragalactic radio sources due to scattering in the lens
\citep{1996ApJ...470L..23J, 1999MNRAS.305...15M, Biggs:2004fw,
  2003ApJ...587...80W}, although in many cases the lens galaxy is
confirmed or suspected to be of late-type. In fact, we
know very little about the turbulent density fluctuations $\langle
\delta n_{\rm e}^2\rangle^{1/2}/\langle n_{\rm  e}\rangle$ in giant
elliptical galaxies (or in general any galaxies other than our own),
so it is unclear whether the scattering measure is larger or smaller
than the estimate provided here. To be conservative, in the following we take
$\mathrm{SM}= 10^{-3}\,\mathrm{kpc\ m}^{-20/3}$ as our fiducial
value.

In our case, the diffractive length $r_{\rm diff}$ over which the rms phase variation
due to scattering equals to one radian is greater\footnote{In case
  $\mathrm{SM}\gg 10^{-3}\rm\, 
  kpc\, m^{-20/3}$ \citep[e.g. when the light ray happens to pass through 
  some dense HII regions or a spiral arm,][]{1996ApJ...470L..23J,
    2003ApJ...590...26W}, we may have $r_{\rm 
    dif}< l_0$, and then the 
  dependence on frequency will be $r_{\rm diff}\propto \nu$,
  which leads to angular broadening $\theta_{\rm scat} \propto
  \nu^{-2}$ and temporal broadening $\tau \propto \nu^{-4}$, and our
  results on temporal broadening will differ by a factor of a
  few.} than the inner scale
$l_0$, so we have
\citep{1992RSPTA.341..151N, 2013ApJ...776..125M}
\begin{equation}
  \label{eq:6}
r_{\rm diff}\simeq \left(3.2\times10^{9}\ \mathrm{cm} \right)\,
\left(1+z_L \right)^{6/5}\,\nu_9^{6/5} 
  \left({\mathrm{SM}\over 10^{-3}\
      \mathrm{kpc\ m}^{-20/3}}\right)^{-3/5}
\end{equation}
Incoming radio waves are scattered into an angle $\theta_{\rm scat} \simeq
\lambda/[2\,\pi\, r_{\rm diff}\,(1+z_L)]$, and the temporal broadening is
given by
\begin{equation}
  \label{eq:8}
  \tau \simeq {(1+z_L)\,D_{\rm eff}\, \theta_{\rm scat}^2 \over c},
\end{equation}
where $D_{\rm eff} = D_L\,D_{LS}/ D_S$.
Taking $z_L = 0.5$ and $z_S = 1.0$ for the example in \refsec{simulation}, we have $D_{\rm
  eff} = 533\rm\ Mpc$ and the temporal broadening is
\begin{equation}
  \label{eq:9}
  \tau\simeq \left(31\,\mathrm{ms} \right)\,\nu_9^{-4.4}
  \left({\mathrm{SM}\over 
      10^{-3}\ \mathrm{kpc\ m}^{-20/3}}\right)^{6/5}\,\left( \frac{1 + z_L}{1.5} \right)^{-4.4}.
\end{equation}
The angular broadening is given by $(D_{LS} /D_S )\, \theta_{\rm scat} \simeq (0.17\, \mathrm{mas})\, \nu_9^{-2.2}\,\left( 1 + z_L \right)^{-2.2}\,(\mathrm{SM}/10^{-3}\
\mathrm{kpc\ m}^{-20/3})^{3/5}$, which may be resolved
by VLBI at sufficiently low frequencies.

The above analysis suggests that scattering broadening by the lens galaxy is enhanced by the cosmological distance leverage. 
It may strongly limit the accuracy of delay-time measurement, and more importantly,
some of the images may have a fluence too temporally spread out to be detectable at all. However, as can be seen in \refeq{9},
scattering broadening is rapidly suppressed toward higher frequencies. For instance,
observing at $3\rm\,GHz$ instead of $1\rm\ GHz$ reduces temporal
broadening by a factor of $3^{-4.4}\simeq 8\times10^{-3}$ for
Kolmogorov turbulence $\beta = 11/3$, bringing down the temporal
broadening to $\tau \simeq 0.25\,{\rm ms}$. Thus, the
method described in this paper will still be useful at a few GHz,
which is accessible at SKA1 and at many of the VLBI instruments. In
fact, going to higher frequencies decreases not only scattering broadening but also 
intraband dispersion, so pulse time of arrival may be measured to an accuracy
better than $\sim1$ ms. 

Even though accurate timing may not be achievable for survey
telescopes at low frequencies $\lesssim 1\,$GHz (such as CHIME and
UTMOST), they may still be able to detect lensed bursts (albeit
severely broadened). Note that the signal-to-noise 
ratio depends on both the fluence $\mathcal{F}$ and de-dispersed pulse 
width $\tau$ as $\mathrm{S/N}\propto
\mathcal{F}\,\tau^{-1/2}$. Given the fact that many bursts with
duration $\sim 10\rm\ ms$ have been detected
\citep[e.g.][]{2016MNRAS.460L..30C, 2017MNRAS.468.3746C}, 
further broadening by a factor of $\sim 10$ (to $10^2\rm\ ms$) will
decrease $\mathrm{S/N}$ by a factor of $\sim3$ (the fluence is
unchanged). Future telescopes may be a factor of a 
few more sensitive than current ones. Moreover, strongly lensed
images typically have magnification factors of a few. Therefore, it is entirely
possible to at least detect strongly lensed (and temporally broadened) FRBs at
$\lesssim 1\rm\ GHz$. On the other hand, we strongly encourage carrying out blind FRBs
surveys at higher frequencies ($\gtrsim 3\rm\ GHz$) and increasing the
maximum pulse width within which FRBs are being searched for. Turning the argument around, abnormally large burst width may be a hint for an intervening lens galaxy.

\subsection{Scattering in the host galaxy}
\label{sec:host-scatter}

The assumption of a point source for the computation of galaxy lensing could
be invalidated by significant scattering in the host galaxy. A large fraction
($\sim1/2$) of known FRBs show frequency-dependent asymmetric pulse
broadening, with scattering times at $\sim$ 1 -- 10 ms at 1 GHz
\citep{2016arXiv160505890C}. The scattering is inconsistent with
being due to the Milky Way along the observed lines of sight or due to the IGM
 \citep{2014ApJ...785L..26L, 2015Natur.528..523M,
  2016arXiv160505890C, 2016ApJ...832..199X}. In the strong scattering
regime, scattering in 
the host galaxy by an effective thin screen at a distance $D_{\rm
  h}$ from the source with typical scattering angle $\theta_{\rm sca, 
  h}$ gives rise to temporal broadening
\begin{equation}
  \label{eq:7}
  \tau_{\rm h} \simeq {D_{\rm h}\,\theta_{\rm sca, h}^2\over c}
\end{equation}
and a larger source size
\begin{equation}
  \label{eq:5}
  \ell_{S} \simeq D_{\rm h}\,\theta_{\rm sca,h}\simeq \sqrt{D_{\rm
      h}\tau_{\rm h}c}\simeq \left(1\times10^{13}\ \mathrm{cm} \right)\, D_{\rm h,
    pc}^{1/2}\, \tau_{\rm h,ms}^{1/2},
\end{equation}
where $D_{\rm h,pc} = D_{\rm h}/\mathrm{pc}$ and $\tau_{\rm h,ms} =
\tau_{\rm h}/\mathrm{ms}$. For a lensed image with
a deflection angle $\alpha\sim 1''$, the additional temporal broadening due to
lensing is (c.f. \refeq{delTkijs}) 
\begin{equation}
  \label{eq:10}
  \tau_{\rm \ell} 
\sim {\alpha\, \ell_S\over c}\simeq (1.6\ \mathrm{ms})
  \left({\alpha\over 1''} \right)\,D_{\rm h,
    pc}^{1/2}\, \tau_{\rm h,ms}^{1/2}.
\end{equation}
Since $\tau_{\rm h}\propto \nu^{-4}$, the temporal
broadening due to lensing scales as $\tau_{\rm \ell}\propto \nu^{-2}$,
which can potentially be used to probe the location of the scattering
screen (e.g. either near the progenitor $D_{\rm h}\sim\rm pc$, or far in the
ISM $D_{\rm h}\sim \rm kpc$). At sufficiently high frequencies
($\gtrsim 3\rm\ GHz$), both $\tau_{\rm h}$ and 
$\tau_{\rm \ell}$ become negligible compared to the intrinsic
width, and the method proposed in earlier sections is still applicable.

Dense ISM in the host galaxy might cause large but coherent refraction, which deflect the propagation of radio waves. Due to relative peculiar motions between the source, the lens, and the Earth, the light ray samples the inhomogeneous distribution of free electrons. As a result, from the perspective of the lens, the apparent location of the emission spot can have random shifts transverse to the line of sight. Sufficiently large apparent shifts may imprint stochastic perturbations in the time delay, in a way similar to the scenario of \refsec{rand-spot}.

\section{Conclusion}
\label{sec:concl}

With good prospects for detecting a large number of FRBs at cosmological redshifts using forthcoming radio telescopes, finding strongly lensed sources is not unthinkable. Moreover, it is possible that repetition is a generic feature for FRBs. In that case, a multiply-imaged repeater, if uncovered from the burst catalogue, would enable measurement of time delay to millisecond accuracy.

In this paper, we have worked out how the motions of the Earth, of the
lens galaxy, and of the source generate perturbations to the time
delay for a generic lensing configuration. The orbital motion of the
Earth induces a large sinusoidal modulation to the delay time $\sim
10^3\,$s, which may be used to narrow down source position in the case
of poor sky localization. More interestingly, if VLBI follow-ups
resolve multiple images, time-delay perturbations can be used to probe
non-uniform source motion, hence providing valuable information about
the astrophysical details of the source. For that purpose, the effects of unknown cosmic peculiar motions for the source and the lens are modeled as linear drift in the delay, and then the effect of the Earth's orbital motion is accurately subtracted. Our proposed
method relies entirely on direct observables and does not require
modeling of the lens. 

Using mock observations, we have demonstrated that source orbital motion with a size $\gtrsim 0.1 - 1\,{\rm AU}$ on a timescale $\sim 1\,{\rm yr}$ is measurable, assuming a timing accuracy of $\sim 1\,{\rm ms}$. This is based on a conservative repetition rate $\sim 0.05\,{\rm day}^{-1}$ and only require monitoring repetitions for $1-2$ years. Key orbital parameters such as orbital period and semi-major axis can be recovered. This will reveal the possible existence of a stellar companion if FRBs require a compact star in a special environment. For other FRB mechanisms, source regions may vary across a distance $\gtrsim 1\,{\rm AU}$. Those scenarios will also be constrained by time-delay perturbations. Moreover, refraction by dense materials in the host system may cause apparent shifts in the source location as viewed from the lens, which may also leave noticeable imprints in the lensing delay time.

At low frequencies $\lesssim 1\,$GHz,  scattering broadening to $\gtrsim 30\,$ms by the lens galaxy could degrade timing accuracy. The effect is significantly larger than scattering in the host galaxy and in the Milky Way. If this does not completely prevent detection, large scattering broadening should hint at intervening objects and hence strong lensing event. Therefore, extending burst search to larger burst widths can be useful for finding lensed FRBs. As long as a lensed source is found, scattering broadening should not pose an issue at high frequencies $\gtrsim 3\,$GHz. 

Finally, we note that, even with scattering broadening, timing accuracy for FRBs much better than $\sim 1\,{\rm ms}$ may be possible through the technique of de-scattering the voltage timestream (i.e. the voltage signal as a function of time at the receiver) using bright bursts \citep{Pen:2014jqa, Main:2017ghn}, if bursts are intrinsically very narrow. In that case, non-trivial source motion on scales smaller than $0.1 - 1\,{\rm AU}$ may be probed. At such a high degree of timing accuracy, further study is needed to see if the time delay models introduced here are sufficient.


\acknowledgments

The authors thank Ue-Li Pen, Dmitri Uzdensky and Siyao Xu for useful feedbacks. We are especially thankful to Timothy Brandt for his useful suggestion on parameter estimation and for carefully reading through an earlier version of the draft. The authors acknowledge the hospitality of the Aspen Center for Physics where this work was initiated. 
LD is supported at the Institute for Advanced Study by NASA through
Einstein Postdoctoral Fellowship grant number PF5-160135 awarded by
the Chandra X-ray Center, which is operated by the Smithsonian
Astrophysical Observatory for NASA under contract NAS8-03060. WL is
supported by the Named Continuing Fellowship at the University of
Texas at Austin.


\bibliographystyle{aasjournal}
\bibliography{reference_frbtimedelay}



\end{document}